\def\a{\alpha}
\def\b{\beta}

\def\d{\delta}
\def\e{\epsilon}
\def\f{\phi}
\def\g{\raisebox{.4ex}{$\gamma$}}

\def\k{\kappa}
\def\l{\lambda}
\def\m{\mu}
\def\n{\nu}

\def\r{\rho}
\def\s{\sigma}
\def\t{\tau}

\def\x{\xi}

\def\D{\Delta}
\def\F{\Phi}
\def\G{\Gamma}

\def\ha{{1\over 2}}

\newcommand{\vx}{\vec{x}}

\def\ltap{\ \raise.3ex\hbox{$<$\kern-.75em\lower1ex\hbox{$\sim$}}\ }
\def\gtap{\ \raise.3ex\hbox{$>$\kern-.75em\lower1ex\hbox{$\sim$}}\ }

\def\gl{\ \raise.5ex\hbox{$>$}\kern-.8em\lower.5ex\hbox{$<$}\ }
\def\roughly#1{\raise.3ex\hbox{$#1$\kern-.75em\lower1ex\hbox{$\sim$}}}

\def\be{\begin{equation}}
\def\beq{\begin{equation}}
\def\te{\end{equation}}
\def\ee{\end{equation}}
\def\eeq{\end{equation}}

\def\bea{\begin{eqnarray}}
\def\ba{\begin{eqnarray}}
\def\nn{\nonumber}
\def\tea{\end{eqnarray}}
\def\eea{\end{eqnarray}}
\def\ta{\end{eqnarray}}

\newskip\humongous \humongous=0pt plus 1000pt minus 1000pt

\newif\ifdtup

 \def\(#1){(\ref{#1})}

\makeatletter                    
\@addtoreset{equation}{section}  
\makeatother                     

\documentclass[preprint,floatfix,nofootinbib,nobibnotes,aps,12pt]{revtex4-1}
\usepackage{amssymb,amsmath,amsthm,graphicx,color}

\begin{document}

\title{Infrared Behavior of Quantum Fields in Inflationary Cosmology \\ -- Issues and Approaches: an  overview}

\author{B. L. Hu}
\affiliation{Maryland Center for Fundamental Physics and Joint Quantum
Institute,\\ University of Maryland, College Park, Maryland 20742-4111, USA}
\email{blhu@umd.edu}
\date{December 26, 2018}

\begin{abstract}
This is a  pedagogical guide to works on this subject which began in the 80’s but has seen vibrant activities in the last decade. It aims to help orient  readers, especially students, who wish to enter into research but  bewildered by the vast and diverse literature  on this subject. We describe the three main veins of activities: the Euclidean zero-mode dominance, the Lorentzian interacting quantum field theory and the classical stochastic field theory approaches in some detail, explaining the underlying physics and  the technicalities of each.  We show how these approaches are interconnected, and highlight recent papers which contain germs of worthy directions for future developments.  
\end{abstract}


\baselineskip=18pt

\allowdisplaybreaks

\maketitle

\noindent{\bf Preface}

Infrared physics means physics at large scales. For quantum field theory in flat space, infrared divergences are often treated by introducing a low frequency cutoff and casting away terms containing this cutoff,  with the rationale that the IR cutoff scale is far away from the region of interest.  This prescription of convenience is obviously inapplicable in at least three circumstances: i) flat space with boundaries, where the boundary conditions  bear real physical consequences;  ii) curved spacetimes where the large scale structure of spacetime, including nontrivial topology, can play an essential role in the IR physics, iii) critical phenomena, where the correlation functions go to infinity at critical points. Systems confined in a finite region show IR behavior different from that in the bulk.  
We are interested in ii) in relation to symmetry breaking in curved spacetime, and as we shall see, the physics of iii) enters in an essential way. What operates there can help us understand phase transitions in the early universe, in particular,  inflationary cosmology. 
IR behavior of quantum fields in curved spacetime is both a mathematical problem and a physical issue: Mathematically, what are the field-theoretical techniques which can properly handle IR divergences? The dominant IR contributions from quantum field modes at superhorizon scales constitute a nonperturbative effect which defies loop-expansion calculations. Physically, the late time behavior of quantum fields under inflation directly affects the physical processes in the post-inflationary eras and determines the cosmological parameters which enter into the later stages where observational data are taken. Questions related to the IR issue we can ask are, e.g., i) Are there secular effects of quantum fluctuations which persist till late times, ii) If yes, will they backreact on the cosmological constant and attenuate its value? iii) Will the IR behavior of quantum fields render the de Sitter universe unstable? These are important questions to ask, but only a valid calculation based on sound techniques can provide trustworthy physical answers.\\

\noindent{\bf Notes} 

{\it The Emphasis}  here is more on pedagogical methods than on review coverage, in the sense that we may attribute somewhat different weights  to equally important papers based on their pedagogical value in helping to clarify the main issues.  Apologies to authors whose works are  not discussed at length here. When we say this is a pedagogical guide, we don’t mean pedagogical in terms of giving full derivations of all the formulas or even the important ones, which the reader can find from the  original papers. We mean pedagogical in the sense of helping the reader to see the main veins of development and highlighting the methods used. Thus this is more in the nature of a guide than a review. We also pay attention to papers attempting to connect these three major approaches, so we can see the physics they share in common, each shedding a different light on it.  

{\it The Source} of this overview is a chapter in the forthcoming book {\it Semiclassical and Stochastic Gravity -- Quantum Field Effects on Curved Spacetime} by  Bei-Lok B. Hu and Enric Verdaguer, Cambridge University Press 2019.  The reader will see references to discussions in earlier and later chapters.  To  make it more or less self-contained we have added some background materials, when necessary. In places where equations of earlier or later chapters are referred to, we  direct the readers to the appropriate equations in the original papers.

\section{Relevance, Issues and Approaches}  

In Chapters  2, 4, 5  we have discussed the ultraviolet (UV) divergences arising from the contributions of  high frequency or short wavelength modes in quantum field theories (QFT) in curved spacetimes (CST) and introduced three methods -- dimensional, zeta-function and point-separation -- to regularize them. In this chapter we focus on the opposite domain, namely,  infrared (IR) divergences arising from the long wavelength modes in the spectrum of the fluctuation wave operator. Traditionally,  in flat space QFT,  IR divergences used to be fixed in a rather simplistic way,  by introducing a low  frequency cutoff in the integration over modes, and  terms containing such cutoffs are cast away, with the justification that they represent processes at far distances beyond the realm of concern to the problem under study.  This kind of argument obviously fails for field effects even in flat space with boundaries. There is nontrivial physics associated with the  boundary and different boundary conditions imposed on the field also make a difference.  (This is an understatement  in the era when AdS-CFT correspondences show up in many areas of physics.)  In the same vein, in curved spacetimes,  the large scale structure of spacetime is expected to play a nontrivial role in the IR physics,  not to mention the need for covariance, where, like the regularization of UV divergences, cut-off function methods are often untenable.  The curvature and topology of spacetime, and how the field is coupled to the spacetime,  become important factors in the investigation of IR physics. 

There was significant progress in the 70's with the invention and application of quantum field theory techniques to the treatment of infrared problems, notably in the works of  Coleman, Cornwall, Jackiw, Politzer, Tomboulis \cite{Col73,CJP74,CJT74,Jac74} in the two-particle irreducible effective action, the  large N expansion of `tHooft \cite{tHo74}  and the decoupling theorem of  Appelquist \& Carazzone \cite{AppCar75}.  A well-known example is the treatment of divergences of QFT at high temperatures by Dolan \& Jackiw \cite{DolJac74} and Bernard \cite{Ber74}.  

Infrared physics is at  the focus of  critical phenomena,  an important area of physical, biological,  even social sciences containing many profound ideas.  Near the critical points the correlation functions often show divergences.  To understand  how a system behaves physically near a critical point,  one needs to calculate the critical exponents in order to place the system in the correct universality class.  Significant progress took place also in the 70's,  with the invention and application of renormalization group techniques in the seminal works of Wilson, Fisher, Kadanoff, Ma, Brezin, Jinn-Justin and many others,  the description of which can be found in excellent monographs, e.g., \cite{JJCriPhe96}.  Of particular relevance to the themes of this chapter are  the works of  Mermin and Wagner \cite{MerWag66}, Hohenberg \cite{Hoh67}, Fisher \cite{Fisher71}, Brezin and Jinn-Justin \cite{BreZin85}, to name just a few.   For QFT in CST understanding phase transitions in the early universe in  fact is an important  driving force toward detailed studies of the infrared behavior of quantum fields in curved spacetime, especially so after the advent of inflationary cosmology. 

\subsubsection{Phase Transitions in the Early Universe}

The very early universe can at certain stages become vacuum energy
dominated and undergo inflationary expansions.  The vacuum energy
source can be from Grand Unification interaction (GUT scale $M_{GU}
\sim 10^{14}$ GeV), quantum gravitational (QG)  interactions (Planck  scale, $M_P \sim 10^{19}$ GeV) or from interactions in some
intermediate energy scales (supersymmetry, Kaluza-Klein, etc).
Guth \cite{Gut81} first pointed out how an inflationary stage in the
early history can help to resolve some outstanding problems
(horizon, flatness, fluctuations) in cosmology.  The model he used,  now  called `old'  inflation, to illustrate these ideas is by tunneling, which may fail the mission because of inhomogeneities generated by infrequent bubble collisions.  The model proposed by Albrecht \& Steinhardt \cite{AlbSte82} and Linde \cite{Lin82}, now called `new' inflation,  uses a flat Coleman-Weinberg potential \cite{ColWei73} associated with a massless field, whereby inflation arises as `slow-roll'. As it avoids the many bubble collision challenge -- our universe is anchored in only one bubble -- it offers a nice `graceful exit' scheme, except for the fact that the self-interaction coupling constant of  the field is too weak to be realistically implementable.  Many inflationary models have been proposed in the last three decades since then, aiming at specific desirable features to better match with obervations.  Suffice it to say here that the study of symmetry breaking in curved spacetime and phase transitions  in the early universe --  old inflation being an example of first order, and new inflation, of second order -- are essential for the understanding of early universe cosmology.  

From a physical point of view one can also appreciate why we need to gain a better understanding of IR physics in de Sitter space.  Take the simple case of eternal inflation,  where, in the  flat-RW coordinatization of de Sitter space,  the scale factor $a(t)$ in a FLRW universe goes like $a(t) = e^{Ht}$ with $t$ the cosmic time and $H=\dot a / a$  the Hubble expansion rate,  also defining the de Sitter horizon.  What happens to the field under this inflationary expansion is that the tower of higher  (frequency)  modes collapses rapidly to the lowest mode. The heaping-on of these long wavelength modes generates IR divergence.  This is both a mathematical problem and a physical issue: Do we have the needed field-theoretical techniques to properly handle this IR divergence? As we shall soon see, the dominant IR contribution from these modes at superhorizon scale  is a nonperturbative effect which defies loop expansion calculations. Physically, the late time behavior of quantum fields under inflation directly affects the physical processes in the post-inflationary eras and determines the cosmological parameters which enter into the later stages where observational data are taken.  For example, are there secular effects of quantum fluctuations which persists till  late times, will they backreact on the cosmological constant and attenuate its value? Will the IR behavior of quantum fields  render the de Sitter universe unstable?  These are important questions to ask, but only a valid calculation based on sound techniques can provide trustworthy physical answers.  We can see that this is no longer an academic issue, it begs for actual verifiable solutions.

What then are the relevant issues involved in, and the best methods useful for,  the study of symmetry breaking in curved spacetimes, as applied to phase transitions in the early universe, and in particular,  the infrared problem in dS space for inflationary cosmology?  

\subsubsection{Effective potential for symmetry behavior in curved spacetimes} 

In the discussion of symmetry behavior it is important to know where
and when minimum free energy states (local and global minimum) exist
and how the system chooses between them -- from energy and entropy considerations -- and evolves from one to another -- via tunneling, nucleation, first order or spinodal decomposition, second order transitions.
For these purposes the approach based on the effective action proves most powerful. 
The effective action $\Gamma(\bar{\phi})$ gives the free energy density of the system as a functional of the order-parameter
field $\bar{\phi}$.  From $\Gamma(\bar{\phi})$ one can compute  field-theoretical and thermodynamic quantities of interest in the
system.  Since contributions from the quantum and thermal fluctuations 
are built in, one does not need to solve separately the equations of motion for the background field and the fluctuation field  as in the
effective-mass approach often used in stability analysis. 
which may not satisfy the self-consistency condition in the iterations.


To focus on  the effect of field coupling, curvature and topology on symmetry breaking, we can restrict our attention to static spacetimes like the Einstein universe or Euclidean spacetimes  where the order parameter field is a constant.  
For constant background metrics and background fields, one can work with the effective potential 
$V(\bar{\phi})= -(\mathrm{Vol})^{-1}\Gamma(\phi)\mid _{\phi=\bar \phi}$, where the spacetime four-volume $\mathrm{Vol}$ is factored out from the effective action. 
These effects  were studied methodically by a few forerunners in the 80's, examining the symmetry behavior of  interacting quantum fields, specifically,  in the  Einstein universe, the Taub universe and the Euclidean de Sitter universe by O'Connor and Shen \cite{OCPhD,ShenPhD} in their PhD thesis works.  For a taste of what this entails, recall in  Chapter 4, after we have derived the effective potential obtained by zeta function method we gave a brief discussion in how to obtain the IR behavior of a $\phi^4$ theory in the Einstein universe.  In contrast to identifying the UV divergences in the small curvature limit,  for IR behavior one should consider the high curvature regime (corresponding to small conformally-related effective mass $M^{2}_{1} a^{2}\ll 1$, with $M_1^2$ in \eqref{M1}). 

The effect of the dynamics of spacetime on the symmetry behavior of a quantum field is more difficult since the background (order parameter)
field $\bar{\phi}$ depends on time. One needs to use the effective action, not the effective potential, and include dynamical quantum processes  such as particle production, discussed in Chapters 2 \& 3, which can be  significant if the phase transition occurs near the Planck scale.  We will see some aspects of this in our discussion of  QFT in Lorentzian dS.



Phase transitions in the de Sitter universe have been studied by many  authors.  The effective potential for $\lambda \Phi^4$, scalar quantum electrodynamics (SQED) and SU(5) gauge fields have been worked out earlier in the 80's and 90's, but the understanding of the infrared behavior of these theories was incomplete.  Here we will focus on the IR problem exclusively.

\subsubsection{Three veins of developments and four parts in this chapter}

This chapter consists of 4 main parts. The first part explains in  general terms how to identify the dominant contributions to the infrared (IR)  behavior of interacting quantum fields in a compact space.   If a band gap exists in the spectrum of the wave equation fluctuations operator  one can identify the zero-mode or the lowest band as giving the dominant contributions to the IR behavior.  Hu and O'Connor (H\&O) \cite{HuOC87} first pointed this out for quantum fields in curved spacetimes and derived the $\sqrt{ \lambda}$ behavior of the (dynamically- generated)  effective  mass  \eqref{IRPS7} from  the curvature of the effective potential  at a minimum energy state  for an  $O(N) \lambda\Phi^4$ self-interacting scalar field  in Euclidean de Sitter space in the leading order large $N$ approximation.   This is a nonpertubative result for the leading IR behavior in Euclidean dS which cannot  be obtained from perturbation theories via loop expansion.   From the IR behavior of scalar quantum electrodynamics (SQED) they also  drew implications for Planck scale inflations \cite{HuOC86}.   With this,  H\&O discussed the ensuing   dimensional reduction in the infrared domain and introduced the notion of effective IR  dimension  (EIRD). They also  posited that the concept of finite size effect in the description of critical phenomena of condensed matter systems captures the essence of the problem of symmetry breaking in curved spacetimes.  We shall follow their exposition here, with examples in some  familiar curved spacetimes.  The second part of this chapter introduces some quantum field theory techniques for treating IR problems. This includes the closed time path (CTP) or `in-in' method developed in Chapters 3 and  the two-particle- irreducible (2PI)  effective action under large N expansion. We shall use the $O(N) \lambda \Phi^4$ field theory as an example to introduce these techniques.  A more formal and systematic presentation of  these  advanced techniques will be presented in the next chapter.  

Having shown a way to treat the IR behavior in Euclidean dS (EdS) --  a 4-sphere $S^4$ in $5D$ Euclidean space $E^5$ -- we turn in the third part to the Lorentzian field theories in dS (LdS) -- a spatially flat ($k=0$) RW coordinatization of dS, sometimes called the `Poincare patch' --   with a dynamical description which is  more often used in the discussion of physical processes in inflationary cosmology.  But  nonperturbative methods in Lorentzian quantum field theories are not so easily accessible (we will mention one method at the end, the nonperturbative renormalization group method).    
One has to rely on resummation techniques order by order, often resulting in rather complicated expressions where the physics is not so easily intrepretable (we will give an example of how this can be done with some clarity).  Here,  in the Lorentzian framework,  the stochastic approach of  Starobinsky and Yokayama (S\&Y) \cite{StaYok94} comes to the rescue. Based on the stochastic inflationary model (SdS) of Starobinsky \cite{Sta86}, the long wavelength sector is treated not as  quantum fields but as classical stochastic variables -- they become classical after decoherence and they become  stochastic when  driven by a noise source representing the effects of the short wavelength modes. The  dominant nonperturbative IR behavior manifests in the correlation functions of stochastic field variables which can be evaluated by distributional averages rather than expectation values as for quantum field operators.  

It is of interest to see how the symmetry behavior calculated in a Lorentzian (RW coordinate) de Sitter space compares with that calculated in the Euclidean coordinates.  The time scale associated with exponential expansion, $1/H$ corresponds to the length scale of the longest wavelength, namely, the radius of $S^4$ in the Euclidean coordinatization of de Sitter universe. In this Lorenzian QFT rendition, the IR physics arises from what H\&O called the dynamical finite size effect \cite{HuEdmonton}. The `finite size' in Lorentzian picture  particular to eternal inflation is dynamically generated: under continous exponential expansion an observer would see physical processes as if they were static  (`frozen' in time).  The zero-mode  in Euclidean dS corresponds to the longest wavelength mode in the Lorentzian dS.  The zero-mode dominance of IR behavior in EdS has the correspondence in LdS:   It is the late time behavior of these superhorizon modes which encode the IR behavior of the  system. In this light we can see the Euclidean picture drawn by  H\&O and the Lorenzian stochastic picture of S\&Y are equivalent. 

These two key ideas in treating IR behavior of quantum fields in curved spacetimes, the Euclidean zero-mode proposed in the 80's  and the stochastic approach proposed in the 90's laid dormant until much later,  in 2005, when Tsamis and Woodard \cite{TsaWoo05} discovered the  potency of S\&Y stochastic approach for addressing  IR  issues,  and in 2010, when  Rajaraman \cite{Raj10} discovered the   zero-mode dominance of IR in Euclidean dS,  as known and used earlier in H\&O's work. From 2008 onward we see an upsurge of activities on IR in dS  issues.  (For a review of this topic up to 2010,  see \cite{See10}.) Much of the developments in the IR in dS problem since around 2005  
can be sorted out following these {\it three main veins} -- the Euclidean, the Lorentzian QFT and the stochastic approach.  In the second subsection we will  provide a guide to the important recent work along these three main lines of development and the connections built around them, where further proofs of the equivalence of the IR results from the Euclidean and Lorentzian QFT, and stochastic approaches are provided. 


In the last part of this chapter we highlight three topics, two are from the most recent works (in 2018): 1) a resummation method \cite{LMT18} for taking into account the higher modes' influence in Lorentzian dS QFT over and beyond the zero mode dominance in Euclidean dS;  2)  the nonperturbative renormalization group method \cite{GauSer15,GuiSer17} which demonstrates  curvature-induced effects, effective IR dimensional reduction, several of the salient features covered in this chapter,  and through it,  a proof of the stability of dS against infrared quantum fluctuations  \cite{MorSer18};  3) IR behavior of gravitons and the gauge issue,  graviton loop contributions and nonlinear effects.   These latter topics will be discussed in more details in the last four chapters of this book.

We begin with a short calculation showing how the IR divergences appear  in a free massless minimally-coupled scalar field in the de Sitter spacetime.

 


\subsection{Infrared divergences in massless minimally-coupled free scalar field}

To see where the IR divergence appears we follow the exposition of Ford and Parker \cite{ForPar77}.  
Consider a massless minimally-coupled scalar field $\Phi(\mathbf{x},t)$ obeying the Klein-Gordon equation 
\be
\Box \Phi (\mathbf{x},t) = 0 
\ee in a RW spacetime with metric 
\be  ds^2 =  -dt^2 + a^2(t)\,d\mathbf{x}^2 .
\ee  
Decomposing $\Phi (\mathbf{x}, t) $ into the spatial and temporal parts  
\be 
	\Phi(\vx, \t) = \int\!\!\frac{d^3k}{(2\pi)^{3/2}} \;\Bigl[ A_{\mathbf{k}}\, e^{i\,\mathbf{k} \cdot \mathbf{x}} \phi_k (\t) + \mathrm{H. C.} \Bigr]   \label{FP(3.4)}
\ee   
(Here $k = \lvert\mathbf{k}\rvert$ and $\mathrm{H. C.}$  denotes the Hermitian conjugate of the preceding term).   
In  time $\tau \equiv  \int^t dt'/ a^3(t')  $   chosen to simplify the form of the wave equation, the mode amplitude function $\phi_k(\t)$ obeys
\be  
	\frac{d^2 \phi_k(\t)}{d\tau^2}  + k^2 a^4 \phi_k (\t) = 0  \label{FP(3.5)} 
\ee 
Ford and Parker treated two classes of  evolution:  i)  power law  $ a(t) = \s t^c$ and   ii) de Sitter in the $k=0$ RW coordinate  $a(t)=\s e^{Ht}$.  

(i) The \underline{power law} family is a solution of the Einstein equation with equation of state $p = \g \r $ with $\g= (2-3c)/3c $.   The general solution of  ($\ref{FP(3.5)}$) is given by the Hankel functions 
\be 
	\phi_k(\t) = c_1 \t^{\ha} H_\n^{(1)} (k |b|^{-1} a_0^2 \t^b) + c_2 \t^{\ha} H_\n^{(2)} (k |b|^{-1} a_0^2 \t^b), 
\label{FP(3.15)} 
\ee
where $c_1(k)$, $c_2(k)$ are complex numbers and $b= (1-c)/(1-3c)$,  $\nu = (2 |b|)^{-1}$.

Now consider a typical class of correlation functions  
\be \frac{\langle 0  \vert \Phi(x) \Phi(x')  \vert0 \rangle }{  \langle 0 \vert0 \rangle}
\ee  
where $|0 \rangle $ is the state defined by  $A_{\mathbf{k}}\lvert0 \rangle = 0$ for all $\mathbf{k}$.   From (\ref {FP(3.4)}) one gets
\be\label{FP(3.22)}
 \frac{\langle 0  | \Phi(x) \Phi(x')  |0 \rangle }{\langle 0 |0 \rangle}=  \int\!\frac{d^3k}{(2\pi)^{3}}\;e^{i \mathbf{k} \cdot (\mathbf{x} -
\mathbf{x}')}  \phi_k (\t) \phi_k^*(\t').      
\ee

From  the general solution for  $\phi_k(\t)$  in ($\ref {FP(3.15)}$),  using  the asymptotic expansion of the Hankel functions  for $ z \rightarrow 0$
\be
H_\n^{(1)}(z) \approx  -H_\n^{(2)} (z) \approx  - \frac{i}{\pi} \G(\nu) (\frac z2)^{-\n}  	
\ee
the integrand of ($\ref {FP(3.22)}$) is seen to carry the factor $|c_1-c_2|^2 k^{-2\n}$.
Thus if $|c_1-c_2|$ is non-zero as $k \rightarrow 0$ the two- point function will diverge if $\n \geq 3/2$, corresponding to the range of $ c \geq 2/3$ for  power law expansion.  For $c$ in this range, the two-point function will diverge if $c_1(k)$ and $c_2(k)$  are regular near $k=0$, otherwise the Wronskian condition which ensures the unitarity of the field evolution will not be satisfied.  Recall that each possible choice of  $c_1(k)$ and $ c_2(k) $ defines a choice of vacuum state.  

However, as noted by Ford and Parker, it is possible to find choices of $c_1(k)$ and $c_2(k)$  which are singular at $k=0$, and for which  $|c_1-c_2| \rightarrow 0$ as $k \rightarrow  0$. Thus,  in contrast to the situation in flat space, here the well-behaved states correspond to singular coefficients, whereas any regular pair of coefficients yields a state with infrared divergence.  

In examining the energy density they found that the infrared behavior in states with $c_1,  c_2$  regular in a metric with power law expansion
($\frac 23  \le c < 1$, $1 <c \le 2$ )  is such as to prevent these metrics from being self-consistent solutions of  the Einstein equation, and thus should be excluded.

ii) For \underline{exponential expansion}    the mode amplitude functions have the form ($\ref{FP(3.15)}$)  with $\nu = 3/2$ and $b=1/3$.  Again it is seen that choices of $c_1(k)$ and $c_2(k) $ which are regular lead to divergences in the two point functions.


The power law and the exponential cases can be considered together. Using the Hubble constant $H$ and the deceleration parameter $\e$  (the  `slow-roll' parameter in inflationary cosmology terminology)  
\be \qquad H(t) \equiv \frac{\dot{a}}{a} \;\; , \;\;
\epsilon(t) \equiv -\frac{\dot{H}}{H^2} , \ee
the FLRW power law  $a(t) =\s t^c$ and the eternal inflation $a(t)= \s e^{Ht}$ solutions  correspond to $\e = 1/c$ and  0  respectively. The mode amplitude function  for $D$ spacetime dimension in cosmic time $t$ is given by  \cite{MMTW14}
\begin{equation} \phi_k(t) = a^{-(\frac{D-1}2)}
\sqrt{\frac{\pi}{4 (1 \!-\! \epsilon) H}} \, H^{(1)}_{\nu}\Bigl(
\frac{-k}{(1 \!-\! \epsilon) H a}\Bigr) \qquad {\rm with} \qquad \nu
= \frac{D \!-\! 1 \!-\! \epsilon}{2 (1 \!-\! \epsilon)} \; .
\label{epsmodes}
\end{equation} 
We note that for eternal inflation $\e = 0$ the mode function for \textit{massive} scalar field in open de Sitter coordinates  has $\nu =
\sqrt{(\frac{D-1}2)^2 - \frac{M^2}{H^2}}$ \footnote{The closed dS coordinate mode functions  \cite{MarMor09} can be derived by analytically continuing from Euclidean de Sitter \cite{Hig87}}.

From the small $k$ form of the mode functions (\ref{epsmodes}), one can see that this two point function 
\begin{equation}
\phi_k(t) \phi_k^*(t') \rightarrow \frac{4^{|\nu|} (1 \!-\! \epsilon)^{|2\nu|} \Gamma^2(|\nu|)}{4 \pi (1 \!-\! \epsilon) 
\sqrt{H a^{D-1} H' {a'}^{D-1}} } \times \Bigl( \frac{ H a H' a'}{k^2}\Bigr)^{|\nu|} \Bigl[1 + O(k^2) \Bigr] \; 
\label{smallk}
\end{equation} 
suffers from infrared divergences throughout the range from $\epsilon = 0 $ (de Sitter) to $\epsilon = \frac32,  \nu= \frac 23$ (pressure free dust).

The IR problem in inflationary cosmology is more aggravated owing to the rapid red-shifting of the spectrum towards the IR. Tsamis and Woodard \cite{TsaWoo09} showed that infrared logarithm appears in pure quantum gravity, in gravity + fermions, in full scalar-driven inflation, in the scalar sector of scalar-driven inflation, in $\phi^4$ theory, in scalar QED and in Yukawa theory.  No matter how small the coupling constant in the theory is, the continued growth of the scale factor $a(t)$ over a prolonged period of inflation must eventually result in this 1-loop correction becoming order unity. Because higher
loop corrections also become order unity at about the same time, perturbation theory breaks down and that one must employ a nonperturbative technique to evolve further. The eventual breakdown of perturbation theory is a general feature of quantum field theories that exhibit infrared logarithms.

Thus, the question is, what are we to make of this IR divergence in de Sitter universe? Is it a nuisance to be rid of by, say, introducing a low frequency cutoff (which will artificially introduce a zero mode even if there isn't one), as was customarily done in earlier times, or does it contain some real physical meaning? 
Unlike  in flat spacetime  the IR divergences in QFT in curved spacetime are more complicated and they reflect real physical  conditions.  The IR behavior of quantum fields is determined by and also reflects the large scale structure of spacetime.  This is the main theme of this chapter. 


\subsection{Issues, Approaches and Methodology }

\subsubsection{Key Issues}

From this simple example already two of the many factors involved in the infrared problem are apparent: 1) choice of the vacuum, 2) the dynamics. We may add a third factor,  3)  coordinate representations.  The commonly studied ones in de Sitter space are a) the global de Sitter, in the closed ($k=1$) Robertson-Walker (RW) coordinate,  which covers the whole space, b) the $a (t)= e^{Ht} $  in the flat ($k=0$)  RW  coordinate (Poincare patch)  which covers half of the  space, and c) $S^4$ in the Euclidean coordinates of $E^5$.   We may at times refer to case b) as  Lorentzian dS (LdS) and case c) Euclidean dS (EdS). The Euclidean vacuum has been shown to be equivalent to the Bunch-Davies vacuum which is de Sitter invariant. (For more general cases, see \cite{All85,AllFol87,KirGar93}.) The LdS (case b) has attracted greater attention after the advent of inflationary cosmology in the 80's. We shall be discussing IR physics in both cases b) and c) below. 

In view of the rapid surge of activities in recent years on this topic a short guide to the literature may be helpful to the readers who prefer to read the original papers but bewildered by the volume of publications.  As a background,   Ford and Parker \cite{ForPar77} spelled out the infrared divergence issue for scalar fields in dS,  Vilenkin and Ford \cite{VilFor82} and Vilenkin \cite{Vil83NPB} used the superdaisy diagrams to capture the leading IR contributions, a method detailed earlier by Dolan and Jackiw \cite{DolJac74} for treating the IR problem in finite temperature QFT.   Historically and still relevant, {three groups of  papers} are noteworthy, here divided according to the approaches taken: \textbf{(A1)} Hu and O'Connor \cite{HuOC86,HuOC87} (H\&O) treating the self-interacting massless minimally  coupled scalar field in Euclidean dS, first pointed out the IR behavior is dominated by the zero-mode. They also suggested the appropriate  \textbf{(B1}) quantum field theoretical techniques for treating infrared behavior of QFT in CST, namely, the two-particle irreducible (2PI) effective action and the large N expansion. 

In a different approach \textbf{(C1)} in the context of stochastic inflation proposed by  Starobinsky \cite{Sta86} where a Langevin equation   with the short wavelength modes acting as a white noise drives the long wavelength modes treated classically,  Starobinsky and Yokoyama (S\&Y) \cite{StaYok94} calculated the correlation function  in flat ($k=0$) RW- dS arising from the longest wavelength modes at late times, and produced correctly the leading infrared logarithms of scalar potential models at arbitrary loop order.  Tsamis and Woodard \cite{TsaWoo05} who have long been studying two-loop perturbative quantum gravity effects in curved space exclaimed that the S\&Y results `miraculously' captures the leading order graviton loop contributions in the IR.  
This  discovery opened the dialogue between the S\&Y  stochastic approach and its relation to the QFT methods. 
\subsubsection{Approaches and Methodologies}

\noindent {\bf A}. From  zero-mode IR dominance in Euclidean de Sitter to nonperturbative  renormalization group (NPRG) :  $A1 \rightarrow A2 \rightarrow A3 \rightarrow  A4  \rightarrow A5$ 

\noindent {\bf B}. Quantum field theory techniques in Lorentian de Sitter:  2PI effective action, Schwinger-Dyson equation: $B1 \rightarrow B2 \rightarrow B3 \rightarrow  B4  \rightarrow B5$ 

\noindent {\bf C}. Stochastic approach and relation to field theory methods:  $C1 \rightarrow C2 \rightarrow C3 \rightarrow  C4  \rightarrow C5$ \\

  For the {Euclidean approach},  in 2010 \textbf{(A2)} Rajaraman \cite{Raj10}  called the attention to the dominance of the zero mode in determining the IR behavior in the massless minimally- coupled $\phi^4$  theory, an observation made earlier in  Hu and O'Connor \cite{HuOC87} who use this fact to define the effective infrared dimension (EIRD) of field theories in curved spacetimes.    \textbf{(A3)} {Beneke and  Moch} 2013 \cite{BenMoc13} showed that solutions based on truncations of the loop expansion can at best be approximate. With the proper exact treatment of the zero mode, the reorganized perturbative expansion is free from infrared divergences. They demonstrated that for  the Schwinger-Dyson equations derived from a two-particle irreducible effective action, the solutions for the two-point functions to leading IR order take the form of free propagators with a dynamical mass.  They also showed that the functional integrals over just the zero mode coincide with the integrals over the probability distribution functions in stochastic inflation.  \textbf{(A4)} Beyond the leading IR approximation, two-point functions for massive scalar fields on Euclidean de Sitter space to all orders in perturbation theory have been investigated by Marolf and Morrison \cite{MarMor10,MarMor11}, where it is found that these are well defined and that in particular, the field correlations exhibit an exponentially decaying behavior for large separations.  Hollands \cite{Hol12} derived the  correlation functions of massless  $\l \phi^{2n}$ fields in a de Sitter invariant state which holds to arbitrary orders in perturbation theory by an analytic continuation from the corresponding objects on the sphere. He establishes that generic correlation functions cannot grow more than polynomially in proper time
for large time-like separations of the points. Also noteworthy are  the proof of the equivalence between  the Euclidean and the in-in formalisms in de Sitter QFT \cite{HigMarMor11a} and the  so-called quantum no-hair theorem \cite{Hol13}.) 
 While addressing the residual gauge issue  Tanaka and Urakawa \cite{TanUra14}  showed that when the Euclidean vacuum is chosen as the initial state, IR regularity and the absence of secular growth are ensured. Note, however,  the warnings issued by  Miao, Mora, Tamis and Woodard \cite{MMTW14} in  translating Euclidean results  back to the Lorentian spacetime via analytic continuation.   
\textbf{(A5)} Guillieux and Serreau \cite{GuiSer15} investigated  scalar field theories in de Sitter space by means of nonperturbative renormalization group techniques \cite{BerTetWet02,Del12}.  They computed the functional flow equation for the effective potential of $O(N)$ theories in the local potential approximation and studied the onset of curvature-induced effects as quantum fluctuations are progressively integrated out from subhorizon to superhorizon scales. This results in a dimensional reduction of the original action to an effective zero-dimensional Euclidean theory. They showed that the latter is equivalent both to the late-time equilibrium state of the stochastic approach of Starobinsky and Yokoyama \cite{StaYok94} and to the effective theory for the zero mode on Euclidean de Sitter space.   Their investigation of dimensional reduction, symmetry restoration and dynamical mass generation form a closure to these issues raised by  Hu and O'Connor \cite{HuOC86,HuOC87} and seals in its relation with the stochastic approach of Starobinsky and Yokoyama \cite{Sta86,StaYok94}. 	We will describe this in a later section
 
In terms of field theory methods and IR issues, we mentioned the extended series of work by  Tsamis and Woodard  and co-workers.    \textbf{(B2)} Serreau \cite{Ser11}  studied the quantum theory of an $O(N)$ scalar field on de Sitter geometry at leading order in a nonperturbative $1/N$ expansion which resums the infinite series of the superdaisy loop diagrams.  He obtained the de Sitter symmetric solutions of the corresponding properly renormalized dynamical field equations and computed the complete effective potential. The  self-interacting field acquires a strictly positive square mass which screens potential infrared divergences. Serreau and Parentani \cite{SerPar13} computed the four-point vertex function in the deep infrared regime for  the $O(N)$ scalar field theory with quartic self-coupling in de Sitter space in the  large-$N$ limit. They found that resummation of an infinite series of perturbative (bubble) diagrams leads to a modified power law which is analogous to the generation of an anomalous dimension in critical phenomena. The high momentum (subhorizon) modes  influence the dynamics of infrared (superhorizon) modes only through a constant renormalization factor. Their calculation provides an explicit example of effective decoupling between high and low energy physics in an expanding space-time.
  \textbf{(B3)}  Gautier and Serreau \cite{GauSer13} showed that the results from the stochastic method are also  obtainable by solving the Schwinger-Dyson equations.  While it was known earlier that the two methods yield the same results for the leading IR-enhanced corrections from the local self-energy (seagull) diagram, from a leading order large $N$ expansion \cite{Ser11}, these authors extended the agreement to nonlocal self-energy diagram.  
Gautier and Serreau  \cite{GauSer15} computed the self-energy of an $O(N)$ symmetric theory at next-to-leading order in a $1/N$ expansion in the regime of superhorizon momenta, and obtained an exact analytical solution of the corresponding Dyson-Schwinger equations for the two-point correlator.  This amounts to resumming the infinite series of nonlocal self-energy insertions, which typically generate spurious infrared and/or secular divergences. The potentially large de Sitter logarithms resum into well-behaved power laws from which one can extract the  field strength and mass renormalization. 

\textbf{(B4)} Stability  of de Sitter spacetime against infrared quantum scalar field fluctuations is shown recently by Moreau  and   Serreau \cite{MorSer18}. They studied  the backreaction of superhorizon fluctuations of a light quantum scalar field on a classical de Sitter geometry by means of the Wilsonian renormalization group. 
 \textbf{(B5)} Finally,  relating Euclidean with Lorentzian dS QFT,  L\'opez Nacir,  Mazzitelli and Trombetta \cite{LMT16} considered an $O(N) \lambda \phi^4$ scalar field  in $D$-dimensional Euclidean de Sitter space and computed the two-point functions including corrections from the interaction of the zero mode with the higher modes.  In  \cite{LMT18} these authors  computed the long wavelength limit of the two-point functions at next-to-leading order in $1/N$ and at leading order in $\lambda$ which involves a further resummation of Feynman diagrams needed when the two-point functions are analytically continued. They prove that, after this extra resummation, up to  $O (\sqrt \lambda,\; 1/N)$,  including the contribution of the bubble chains diagrams changes the behaviour of the two-point functions, which now tend to zero (instead of a constant) as $r \rightarrow \infty$.  
  
For the connection between the stochastic approach and QFT approach
\textbf{(C2)} Riotto and Sloth \cite{RioSlo08} investigated  the infrared behavior of  a $O(N) \phi^4$ theory  in $k=0$ RW- dS by solving the equation for the correlation function
 $C^{++}(x,x)$ and the gap equation derived earlier by  Ramsey and Hu \cite{RamHu97ON} from a close-time-path 2PI effective action under large $N$ approximation.  They  first evaluate $C^{++}(x,x)$ from Starobinsky and Yokoyama's Fokker-Planck equation, supposedly a good approximation for the late time behavior,  insert this into the gap equation and solve for $C^{++}(x,x')$ obtaining a simple analytic form for it.  	\textbf{(C3)} Tsamis and Woodard \cite{TsaWoo09}  generalized Starobinsky's techniques to scalar models which involve fields that do not produce infrared logarithms such as fermions and photons  and the derivative couplings of quantum gravity. \textbf{(C4)} The authors of  \cite{GarRigZhu14,GGRZ15} provided an explicit linkage between these two approaches.  In particular  \cite{GGRZ15} from a QFT calculation truncated at leading IR order fully reproduces, at all loop order, the stochastic correlation functions found by Starobinsky and Yokohama, thus proving the proposed conjecture. 
Finally, \textbf{(C5)}  Venin and Starobinsky  \cite{VenSta15}  derived  non-perturbative analytical expressions for all correlation functions of scalar perturbations in single-field, slow-roll inflation. They alerted enthusiasts of the stochastic approach to take note that  different choices of time variable in the Langevin equation account for different stochastic processes and the correct time variable to work with is the number of e-foldings  $\a = \ln a$ (they call it $\mathcal{N}$) is the correct time to use for stochastic processes which matched with results from QFT.



\section{Euclidean zero-mode, EIRD, 2PI effective action}

In this section we consider an $N$-component self-interacting scalar field $\Phi^a$ 
$(a=1$, $...$, $N)$ on an Euclidean manifold of dimension $D$.  We analyze the spectrum of the fluctuation field wave operator and show that  the zero mode gives the dominant contribution to its infrared behavior. We then introduce the notion of  effective infrared dimension (EIRD) to show dimensional reduction and  explain the physics as corresponding to finite size effect in condensed matter physics. Following this we provide a more rigorous analysis  in terms of the two-particle-irreducible (2PI) effective action under a large $N$ expansion of interacting $O(N)$ fields in curved spacetime. Our presentation here follows \cite{HuOC87}.

\subsection{$O(N)$ self-interacting scalar field in curved spacetime} \label{sub-IRDSN}

Let us first consider a massive $(m)$ self-interacting $(\l)$ single component scalar field
$\F$ coupled $(\x)$ to a general $n$-dimensional curved spacetime with metric
$g_{\m\n}$ of Lorentzian signature $(-,+,+,+, \cdots)$ and scalar curvature $R$.  It is described by the
Lagrangian density
\be  L_{\f}^{(0)}[\F,g_{\m\n}]=  -\frac{1}{2}\,\F\Bigl[-\square+(1-\x)\x_{{n}}R+m^{2}\Bigr]\F-{\l \over 4!}\,\F^{4}, \label{ndu} \te  
where 
\begin{equation}
\Box \equiv g^{\mu \nu }\nabla _{\mu }\nabla _{\nu }=\frac{1}{\sqrt{-g}%
}\frac{\partial }{\partial x^{\mu }}\left(\sqrt{-g}\, g^{\mu \nu }\,\frac{%
\partial }{\partial x^{\nu }}\right)
\label{LapBel}
\end{equation}
 is the Laplace-Beltrami operator in a spacetime with metric $g_{\mu\nu}$.  It is a generalization of the d'Alambertian wave operator in Minkowski space to curved spacetime. Note here we use $\x=0$, $1$ to denote conformal and minimal coupling, respectively and   $\x_{{n}}={1 \over 4}{({n}-2) \over ({n}-1)}$ is the conformal coefficient in $n$ dimensions.
  
The action has a minimum at $\F=\bar{\f}$, which satisfies the classical equation of motion 
\be A_{2}\,\bar{\f}(x)=\bigl(-\square+M_{2}^2\bigr)\,\bar{\f}(x)=0 \te 
where
\be M_{2}^2=m^{2}+(1-\x)\x_{{n}}R+{\l \over 6}\,\bar{\f}^{2}. 
\label{M2}
\te
Fluctuations $\varphi =\F-{\bar\f}$ around the classical background
${\bar\f}$ satisfy an equation (to lowest order) \be
A_{1}\,\varphi(x)=\bigl(-\square+M_{1}^2\bigr)\,\varphi(x)=0 \label{HCS2.4}\te where \be  
M_{1}^2=m^{2}+(1-\x)\x_{{n}}R+{\l \over 2}\,\bar{\f}^{2}.
\label{M1}\te
We shall also find it convenient to define  $M^2_{\xi} =m^2+(1-\xi)\xi_n R$.  The $M_1$, $M_2, M_\xi$ are  effective masses which depend on the
coupling $\x$, background curvature $R$, and the background field
$\bar{\f}$. 
Contributions of the fluctuation field to the
equation of motion for ${\f}$ enter through the vacuum
expectation value and the thermal average of its variance
$\ha \l\,\langle\f^{2}\rangle$, which acts as additional terms in the
effective mass $M_{2}$. 

Now consider an $N$-component self-interacting scalar field $\Phi^a$ $(a=1$, $...$, $N)$ on an {\it Euclidean} manifold of dimension $D$ (where the signature of the metric is   $(+,+,+,\cdots)$, obtained by changing  the Lorentzian time $t$  to the Euclidean time $\tau=it$) coupled to the background spacetime with scalar curvature $R$ described by the action
\begin{equation}
	S_{\textsc{e}}[\Phi] =i \int d^Dx\sqrt{g} \left[- \frac{1}{2} \Phi^a \Box_{\textsc{e}} \Phi^a
+ \frac{1}{2}M_\xi^2 \Phi^2 + \frac{\lambda}{4!} \Phi^4 \right]
\; , \label{IRON1}
\end{equation}
In the integral over the $D$-dim Euclidean space,  $x^{0}=\tau$ and  $\Box_{\textsc{e}} =\sqrt{g}\, \partial_\mu (\sqrt{g}\, g^{\mu \nu}\partial_\nu) $ is now an elliptic operator.  We shall keep $D$ general here until later when we consider it to be a direct product of a noncompact space of dimension $c$ and a compact space of dimension $b$.  The Euclidean de Sitter space is a $4$-sphere in $E^5$,  thus $c=0, D=b=4$.  In Sec. III, the full space+time dimensionality is denoted as $d$ instead of $D$.

Decompose the field $\Phi^a$
into a background field $\bar{\phi}^a$ and a fluctuation field
$\varphi^a$ , i.e., $\Phi^a = \bar{\phi}^a + \varphi^a$.  The
background field $\bar{\phi}^a$ is required to satisfy the
classical equations of motion with an arbitrary external source.
Such a shift eliminates the linear term in the fluctuation field,
which is equivalent to performing a Legendre transform.  The
resultant action is
\begin{align}
	S_{\textsc{e}}[\bar{\phi},\varphi]&=S_{\textsc{e}}[\bar{\phi}]+\int d^Dx\sqrt{g}\notag\\
	&\qquad\times \left\{\frac{1}{2} \varphi^a \left[\left(-\Box_{\textsc{e}}
+M_\xi^2+\frac{\lambda}{6} \bar{\phi}^2\right)
\delta^{ab}+\frac{\lambda}{3} \bar{\phi}^a \bar{\phi}^b\right]
\varphi^b + \frac{\lambda}{6}\bar{\phi}^a\varphi^a\varphi^2 +
\frac{\lambda}{4!} \varphi^4 \right\}. \label{IRON2}
\end{align}
Notice that there are two kinds of vertices: a three point vertex proportional to $\lambda\bar{\phi}^{a}(x)$ and a four point vertex proportional to $\lambda$.

The effective action $\Gamma_{\textsc{e}}[\bar{\phi}]$ is obtained by
functionally integrating over the fluctuation fields:
\begin{equation}
e^{-\Gamma_{\textsc{e}} [\bar{\phi}]} = \int
[d\varphi]\;e^{-S_{\textsc{e}}[\bar{\phi},\varphi]} .   \label{IRON3}
\end{equation}
where [$d\varphi$] denotes  functional differential of  $\varphi$.
The wave operator $A^{ab}$ for the fluctuating field is given by
\begin{equation}
	A^{ab}= \bigl(-\Box_{\textsc{e}}+M_1^2 \bigr)
\frac{\bar{\phi}^a\bar{\phi}^b}{\bar{\phi}^2} + \bigl(-\Box_{\textsc{e}}+M_2^2\bigr)
\left(\delta^{ab} -
\frac{\bar{\phi}^a\bar{\phi}^b}{\bar{\phi}^2}\right) \label{IRON4}
\end{equation}
Here $\bar{\phi}^a\bar{\phi}^b/\bar{\phi}^2$ and
$\delta^{ab}-\bar{\phi}^a\bar{\phi}^b/\bar{\phi}^2$  are
orthogonal projectors, the former projects along the direction in
the internal space picked out by $\bar{\phi}^a$ and the latter
projects into an $(N-1)$-dimensional subspace orthogonal to the
direction of $\bar{\phi}^a$. Note that the operator $\Box_{\textsc{e}}$ does
not commute with the projectors unless $\bar{\phi}^a$ is a
constant.

When the direction in group space picked out by $\bar{\phi}^a$
does not vary from point to point around the manifold (this does
not necessarily imply $\bar{\phi}^2$ is a constant) 
the Euclidean Green's function for $A^{ab}$ is  given by
\begin{equation}
G^{ab}=G_1 \frac{\bar{\phi}^a \bar{\phi}^b}{\bar{\phi}^2} +G_2
\left( \delta^{ab}-\frac{\bar{\phi}^a\bar{\phi}^b}{\bar{\phi}^2}
\right)\;, \label{IRON5}
\end{equation}
where the $G_i$ ($i=1$, $2$) are the Green's functions for the
operators $-\Box_{\textsc{e}}+M^2_i$, that is, $(-\Box_{\textsc{e}}+M^2_i)G_{i}=-\delta$.

The one-loop contribution is given by the sum of the logarithms of
the determinants of the fluctuation operators.  So in this case,
by using the projection operators, the one-loop effective action
is given by 
\begin{align}
	\Gamma_{\textsc{e}}[\bar{\phi}]&=S_{\textsc{e}}[\bar{\phi}]-\frac{1}{2} {\rm Tr} \ln G_1-\frac{N-1}{2} {\rm Tr}  \ln G_2\notag\\
	&=S_{\textsc{e}}[\bar{\phi}]+\frac{1}{2} \sum_{l}d_l \ln \lambda_{1l} +
\frac{N-1}{2} \sum_{l} d_l \ln \lambda_{2l} \;, \label{IRON7}
\end{align}
where $S_{\textsc{e}}[\bar{\phi}]$ is the classical action, $\lambda_{il}$ and
$d_l$ are the eigenvalues and degeneracies of the operators $-\Box_{\textsc{e}}
+ M^2_i$.




The Euclidean de Sitter spacetime is a 4 sphere  $S^4$ embedded in $E^5$ with volume $\mathrm{Vol}= 8\pi^2a^4/3$.
The 1-loop effective potential for an $N$ component scalar field with $O(N)$ symmetry in de Sitter spacetime is given by
\begin{equation}
V^{(1)} = \frac{1}{2} \sum^{\infty}_{\ell=0} d_\ell \ln(\lambda_{1\ell} \mu^{-2}) + \frac{N-1}{2} \sum^{\infty}_{\ell
 =0} d_\ell \ln (\lambda_{2\ell} \mu^{-2})
\label{IROC1}
\end{equation}
where
\begin{equation}
\lambda_{i\ell } = \frac{\ell(\ell +3)}{a^2} + M^2_i,\qquad i=1,2,\qquad
d_\ell = \frac{(\ell +1)(\ell +2)}{6} (2 \ell + 3). \label{IROC2}
\end{equation}
The summation leads to an ultraviolet divergence which need be regularized. We showed in Chapters 4 and 5 how to apply the zeta-function \cite{DowCri76,Haw77}  and  the point-separation regularization schemes for this purpose.   Here we focus on the infrared behavior.

\subsection{Euclidean zero mode: Effective IR dimension} 

To identify the dominant source of infrared contributions, let us consider 
cases where the eigenvalues of the fluctuation operator takes on a band structure.  By band
structure we mean that the eigenvalues occur in continua with each
continuum having a higher lowest eigenvalue than the previous one.
This is true for fields on spacetimes with compact sections or for
fields with discrete spectrum.  The procedure is to expand the fields in terms of the band eigenfunctions and convert the functional integral over the fields to an integral
over the amplitudes of the individual modes.  When the lowest mode is
massless it will give the dominant contribution to the effective
action.  The low energy behavior corresponds to a lower-dimensional
system.

\subsubsection{Decoupling of the Higher Modes (or Bands)}
\label{sub-IRDC}

On a manifold with topology $R^c \times B^b$ where $B$ is compact,
consider quantum fields where the fluctuation operator $A$  in
(\ref{IRON4}) has the general form of a direct sum of operators
$D$ and $B$
\begin{equation}
A^{ab}(x,y) = D^{ab}(x) + B^{ab}(y)
\label{IRDC1}
\end{equation}
with coordinates $x$ on  $R^c$ and $y$ on $B^b$.  Assume that the
eigenvalues $\omega_n$  associated with the eigenfunctions
$\psi_n(y)$ of $B^{ab}$ are discrete:
\begin{equation}
B^{ab}\psi_n(y)= \omega^{ab}_{n} \psi _n(y).
\label{IRDC2}
\end{equation}
Decomposing the field $\varphi^a(x,y)$ in terms of $\psi_n(y)$
\begin{equation}
\varphi^a(x,y) = \sum_{n} \varphi^{a}_{n}(x)\psi_n(y)
\label{IRDC3}
\end{equation}
one obtains for the quadratic part of the action Eq. \eqref{IRON1} for the $O(N)$ $\lambda \phi^4$ theory,
\begin{equation}
\frac{1}{2} \int\! d^c x d^b y\;\varphi^a A^{ab}\varphi^b = \frac{1}{2}
\int\! d^c x\;\Bigl[\varphi^{a}_{n} f_{nm}D^{ab}\varphi^{b}_{m} +
\omega^{ab}_{n} f_{nm}
\varphi^{a}_{n} \varphi^{b}_{m}\Bigr]    
\label{IRDC4}
\end{equation}
where $f_{nm} = \int d^by\;\psi_n(y)\psi_m(y)$.  When $\varphi_n$
are properly normalized $f_{nm} = \delta_{nm}$ (we will make such
a choice here) the resulting theory in terms of the new fields
$\varphi^{a}_{n}$ will involve massive fields with masses
determined by the eigenvalue matrix $\lambda^{ab}_{in}$ of the
operators $-\Box_{\textsc{e}}+M_i$  
even if the fields in terms of the old variables appeared massless.  
We will take the
smallest eigenvalue to be given by $n=0$ and assume that its only
degeneracy is labeled by the indices $a$ and $b$. Assume also
that the operator $D^{ab}$ is simply minus the Laplacian $\Box_c$
on $R^c$ times $\delta^{ab}$, the $n=0$ mode is then governed by
the action whose quadratic term is
\begin{equation}
\frac{1}{2} \int\! d^c x\;\Bigl[\varphi^a_0(-\Box_c)\delta^{ab}\varphi^b_0 +
\omega^{ab}_0 f_{00} \varphi^a_0\varphi^b_0\Bigr]. \label{IRDC5}
\end{equation}
Thus this appears like an $c$-dimensional field with an apparent
mass matrix  $\omega^{ab}_{0}$. For the case of an $N$ component
$\lambda \phi^4$ theory the action after this decomposition takes
the form
\begin{align}
	S_{\textsc{e}}[\bar{\phi}+\varphi] &= \int\! d^c x\;\Bigl[\frac{1}{2}
\varphi^a_n(-\Box_c \delta^{ab} + \omega^{ab}_n)\varphi^b_n\Bigr.\notag\\
&\qquad\qquad+\Bigl. \frac{\lambda}{6} g^a_{n\ell m} \varphi^{b}_{\ell} \varphi^b_m +
\frac{\lambda}{4!} f_{kn\ell m}\varphi^a_k \varphi^a_n
\varphi^b_\ell
\varphi^b_m\Bigr],   
\label{IRDC6}
\end{align}
where $g^a_{n\ell m} = \int\! d^by\; \bar{\phi}^a \psi_n \psi _\ell
\psi _m$ and $f_{kn\ell m} = \int\! d^by\; \psi_k \psi_n \psi _\ell
\psi _m$.  The effective action is now given by the functional
integral
\begin{equation}
e^{-\Gamma_{\textsc{e}} [\bar{\varphi}]} = \int [d\varphi^a_n]\;e^{-S_{\textsc{e}}[\bar{\phi}
+ \varphi]}. \label{IRDC7}
\end{equation}
The interesting case occurs when the lowest eigenvalue approaches
zero.  At low energy the Appelquist-Carazzone \cite{AppCar75} decoupling theorem
assures us that with higher modes decoupled from the dynamics, the
infrared behavior is governed by the lowest band.  We are then
left with an effective lower dimensional theory.  

\subsubsection{Correlation Length and Effective Infrared Dimension}
\label{sub-IRED}

The above result of dimensional reduction from a formal
derivation of mode decoupling can be understood in a more
physical way by using the concept of effective infrared
dimensions (EIRD).  By effective IR dimension we mean the
dimension of space or spacetime wherein the system at low energy behaves
effectively.  One well-known example is the Kaluza-Klein
theory of unification and the cosmology based on it.  For example, start with an 11-dimensional spacetime with full diffeomorphism symmetry at above the Planck energy scale. After spontaneous compactification it reduces at energy below the
Planck energy to the physical 4-dimensional space with $GL(4,R)$
covariance and a 7-dimensional internal space with symmetry
group containing the standard $SU_3 \times SU_2 \times U_1$
subgroups of strong and electroweak interactions.  For observers
today at  very low energy the effective IR dimension of spacetime
is four, even though the complete theory is eleven dimensional.

For curved-space symmetry breaking
considerations, the EIRD which the system ``feels'' is governed
by a parameter $\eta$ which is the ratio of the correlation
length $\Xi$ and the scale length $L$ of the background space
$\eta \equiv \Xi /L$.  For compact spaces like $S^4$, $L$ is
simply $2\pi$ times the radius of $S^4$.  For product spaces $R^c \times B^b$ with some
compact space $B$, there are two scale lengths: $L_b$ is finite in
the compact dimensions and $L_c = \infty$ in the non-compact
dimensions.  
The symmetry behavior of the system (described here by a $\lambda
\phi^4$ scalar field as example) is determined by the {\em
correlation length} $\Xi$ defined as the inverse of the {\em
effective mass $ M_{\rm eff}$} related to the effective potential
$V_{\rm eff}$ by (we use subscript eff to denote quantities including
higher loop corrections)
\begin{equation}
\Xi^{-2} = \frac{\partial ^2V_{\rm eff}}{\partial \bar{\phi}^2}
\mid_{\bar{\phi}_{min}} \equiv M^2_{\rm eff}  =(^{curvature-induced \;
mass \; M^2 _{1,2}}_{+ \; radiative \; corrections})    
\label{IRED1}
\end{equation}
where $M_{1,2}$ were defined in  \eqref{M1} and \eqref{M2}. 
It measures the curvature of the effective potential at a minimum
energy state $(\bar{\phi} = 0$ for the symmetric state, or the
false vacuum, $\bar{\phi}=\bar{\phi}_{min}$ for the
broken-symmetry state or the true vacuum.)  The effective mass is
defined to include radiative corrections to the same order
corresponding to the effective potential.  (This quantity is
called the generalized susceptibility function in condensed
matter physics.)  The critical point of a system is reached when
$\Xi \rightarrow \infty$ or $ M_{\rm eff} \rightarrow 0.$  In flat or
open spaces or for bulk systems, the critical point can be
reached without restriction from the geometry (note that in
dynamical situations, exponential expansion can effectively
introduce a finite size effect equivalent to event horizons, see
\cite{HuEdmonton}.) 
However, in spaces with compact dimensions,
the correlation length of fluctuations can only extend to
infinity in the remaining non-compact dimensions, and thus the
critical behavior becomes effectively equivalent to a lower
$c$-dimensional system.  

One can also think of $\Xi$ as the Compton
wavelength $\Lambda = 2\pi/M_{\rm eff}$ of a system of
quasi-particles with effective mass $M_{\rm eff}$.  Any fine
structure of the background spacetime with scale $L$ is relevant
only if $\Lambda \leq L$.  Thus when $\Lambda$ is small or $\eta
\ll 1$, (far away from critical point, at higher energy, with
higher mode contributions) it sees the details of a spacetime of
full dimensionality.  At this wavelength, the apparent size of the
universe is large in both compact and non-compact dimensions.
When $\Lambda \rightarrow \infty$ or $\eta \gg 1$ (near critical
point, IR limit, lowest mode dominant) structures of finite sizes
or the compact dimensions will not be so important.  The apparent
size of the universe will be measured by the non-compact
directions and the EIRD is measured by the number of non-compact
dimensions.  The value of $\eta$ getting very large is an
indication of when dimensional reduction can take place.
Notice that in curved-space the coupling parameters also run with curvature or the scale length of the space.  This makes the concept of EIRD even more interesting, as there is now an interplay between $\Xi$ and $L$, and $\eta$ can either decrease or increase with curvature. 

For example, for $\lambda \phi^4$ fields in the Einstein universe near the symmetric state $\bar{\phi}=0$, the EIRD is equal to 1 but near the global minimum of broken symmetry state $\bar{\phi}=\bar{\phi}_{min}$ is equal to 4. Near the symmetric state, $\eta \gg 1$ signifies reduction of EIRD to one. This is consistent with the theorem of Hohenberg \cite{Hoh67}, Mermin and Wagner \cite{MerWag66} (for statistical mechanics on a lattice) and Coleman (for continuum field theory) \cite{Col73} which states that in dimensions less than or equal to two, the infrared divergence of the scalar field is so severe that there could be no possibility of spontaneous symmetry breaking: the only vacuum expectation value for $\bar{\phi}$ allowed is zero.  Away from the region $\bar{\phi} \simeq 0$ the one-dimensional behavior no longer prevails.  Indeed a global minimum of the effective potential exists at $\bar{\phi}_{min}$. 
Near $\bar{\phi}_{min}$, $\eta \ll 1$ and decreases with curvature. Thus the apparent size of the universe near the global minimum actually increases with increasing curvature.  There is therefore no dimensional reduction and the system has a full 4-dimensional IR behavior.  A transition  to the asymmetric ground state is not precluded as symmetry breaking via tunneling is in principle possible.  The complete picture extending from $\bar{\phi}=0$ to $\bar{\phi}=\bar{\phi}_{min}$ is a combination of one dimensional and four dimensional infrared behavior.  Similar arguments can be applied to other spacetimes or field theories.  Using this notion one can understand, for example, why it is often said that at high  temperatures (small radius limit of $S^1$) the finite temperature theory becomes an effective three-dimensional theory.

\subsection{2PI Effective action and infrared behavior} \label{sub-IRON}


In this section we discuss the nonperturbative regime of an $O(N)$ 
quantum field theory via the $1/N$ expansion \cite{BrezinLargeN}.  Many quantum mechanical,
statistical and field-theoretic models representing physical systems are
known to possess internal symmetries.  Several of these theories admit
generalizations in which the number of internal degrees of freedom $N$
parametrizing the symmetry group of the problem may be treated as a free
variable parameter.  The large $N$ expansion looks at the limit when this
parameter becomes very large, but does nothing to restrict the range of the
coupling constants in the theory. Thus it is regarded as nonperturbative in
the coupling constants.

The $1/N$ expansion scheme has proven its efficacy in dealing with a diverse
class of theories ranging from a single-particle potential problems in
quantum mechanics to phase transitions and critical phenomena, and
quantum chromodynamics. (For a collection of representative papers see in e.g., \cite{BrezinLargeN}.) 
Take a flat space quantum field theory example:  the spontaneous  symmetry breaking of a finite temperature system  studied via the one loop effective potential. When expanded in powers  of $\b^{2}/M^{2}$ (here $M$ is an effective mass for the $\l\f^{4}$ theory and  $\b$ is the inverse temperature)  there are terms  nonanalytic in the mass parameter near $M^{2}=0$.  These terms arise from the infrared sector of the system (the $n=0$ mode in the discrete sum and the ${\bf k}=0$ region in the integration) and lead to a complex value for the critical temperature which is unphysical.  
To investigate the nonperturbative regime of the theory Dolan and Jackiw  \cite{DolJac74} considered an $O(N)$ $\l\f^{4}$ theory in the $1/N$ approximation and found that the dominant graphs in the $1/N$ and high temperature expansion are the so-called daisy or cactus diagrams because  for zero mass they are the most infrared singular.  In this way they were able to eliminate the terms that give an imaginary contribution to the
critical temperature, and for weak coupling they found an  acceptable expression for $T_{c}$. The divergence at high temperature of a finite temperature theory exemplies the infrared problem for quantum fields which happens in many contexts. 
 Here we  shall develop the composite operator technique for treating such divergences in a general setting.


To study the nonperturbative properties of a field theory in the
most general setting with nontrivial infrared behavior, we will
use the effective action for composite operators introduced by
Cornwall, Jackiw and Tomboulis \cite{CJT74}.  We will see that using
this method the sum of all daisy type graphs performed by Dolan
and Jackiw can be reduced to the evaluation of only one graph.
The main idea is to construct a generalized effective action from
which one can derive by a variational calculation not only the
full background field but also the dressed two point function.
The relevant object in this case in place of the one particle
irreducible effective action $\G_{\textsc{e}}[\bar{\phi}]$ is the two particle
irreducible (2PI) effective action $\G_{\textsc{e}}[\bar{\phi},G]$ which can be
deduced from the following generating functional \be
e^{-W_{\textsc{e}}[J, K]} = \int [d \Phi]  \mu [\Phi]\;
                                    e^{-S_{\textsc{e}}[\Phi] -\Phi^{a}J_a - \Phi^{a}K_{ab}\Phi^{b}},
\label{IR2P1} \te 
where  $\mu[\Phi]$ is the functional integration measure for the field $\Phi$ and we have used DeWitt's condensed index notation. Here $K_{ab}$ is an external current, and functional
derivatives with respect to $K_{ab}$ generate vacuum expectation
values of an even number of products of ${\Phi}^a$. The
generating functional in $\Phi$ for two particle irreducible
Green functions expressed in terms of the propagator $G$ is given
by performing the Legendre transform of $W$ \be
\G_{\textsc{e}}[\bar{\phi},G]=W_{\textsc{e}}[J,K]-\bar{\phi}^{a}J_{a}-(\bar{\phi}^{a}\bar{\phi}^{b}+G^{ab})K_{ab},
\label{IR2P2} \te where \bea
{\d W_{\textsc{e}} \over \d J_{a}} &=&\langle\Phi^{a}\rangle_{J,K}\equiv \bar{\phi}^{a}\nn\\
{\d W_{\textsc{e}} \over \d K_{ab}} &=& \langle\Phi^{a}\Phi^{b}\rangle_{J,K}\equiv
G^{ab}+\bar{\phi}^{a} \bar{\phi}^{b}.  \label{IR2P3} \tea Solving
these equations with $J_{a}$ and $K_{ab}$ set to zero gives the
true ground state and the connected two-point function on $G^{ab}$
for the field. Substituting for these sources from the equation of
motion one obtains the generating functional \bea
e^{-\G_{\textsc{e}}[\bar{\phi},G]} &=& \int D\Phi\,\mu[\Phi]\,\exp
\left\{-S_{\textsc{e}}[\Phi]+\left(\Phi^{a}-\bar{\phi}^{a} \right)\left({\d \G_{\textsc{e}}
\over \d \bar{\phi}^a}-\bar{\phi}^{b}{\d \G_{\textsc{e}} \over \d G^{ab}}
\right)\right.\nn\\
& &
+\left.\left(\Phi^{a}\Phi^{b}-\bar{\phi}^{a}\bar{\phi}^{b}-G^{ab}\right)
{\d \G_{\textsc{e}} \over \d G^{ab}}\right\} \label{IR2P4} \tea which is an
implicit equation for $\G_{\textsc{e}}[\bar{\phi},G]$ that can be solved
iteratively.  Perform a background field decomposition
$\Phi^{a}= \bar{\phi}^{a}+\varphi^{a}$, where $\varphi^{a}$ is the
fluctuation field,  this  expression simplifies to \be
e^{-\G_{\textsc{e}}[\bar{\phi},G]}=\int D \varphi \, \mu[\bar{\phi}+\varphi]\,\exp
\left\{-S_{\textsc{e}}[\bar{\phi}+\varphi]+\varphi^{a}{\d \G_{\textsc{e}} \over \d
\bar{\phi}^{a}}+\left(\varphi^{a}\varphi^{b}-G^{ab}\right){\d \G_{\textsc{e}} \over \d
G^{ab}} \right\} \label{IR2P5} \te
From this expression it is apparent that $G^{ab}$ is the analog for the two
point function of what $\bar{\phi}^{a}$ is for the one point
function, the net effect being that we replace all internal
propagators by $G^{ab}$.  For the loop expansion we find that \be
\G_{\textsc{e}}[\bar{\phi},G]=S_{\textsc{e}}[\bar{\phi}]+{1 \over 2}\mbox{Tr}\ln[ G^{-1}]
+{1 \over 2}A_{ab}[\bar{\phi}]G^{ab}-{1 \over 2}\mbox{Tr}({\bf
1})+\tilde{\G}_{\textsc{e}}[\bar{\phi},G] \label{IR2P6} \te Here
$\tilde{\G}_{\textsc{e}}[\bar{\phi},G]$ represents the sum of all two particle
irreducible graphs that are present in the ordinary effective
action but evaluated with the propagator $G$, and
$A_{ab}[\bar{\phi}]$ is the small fluctuation operator in the
background field $\bar{\phi}$.

We now apply the above formalism to the $O(N)$ self-interacting
scalar field theory. 
The one loop  effective action is given by (\ref{IRON7}).
 The two loop contribution to the usual effective action which for the
$O(N)$ theory is given by 
\begin{align}
	\G_{\textsc{e}}^{(2)}[\bar{\phi}] &= {\l \over
4!}\int d^{D}x \sqrt{g}\;\Bigl\{\Bigl[\operatorname{Tr}G(x,x)\Bigr]^{2}+
2\operatorname{Tr}\Bigl[G^{2}(x,x)\Bigr]\Bigr\} \nn\\
&\qquad\qquad -{\l^{2} \over 36}\int d^{D}x \sqrt{g}\int d^{D}y
\sqrt{g}\;\;\bar{\phi}^{a}(x) \Bigl\{G^{ab}(x,y)\operatorname{Tr}\Bigl[G^{2}(x,y)\Bigr] \Bigr. \nn\\
&\qquad\qquad\qquad\qquad+\Bigl.2G^{ac}(x,y)G^{cd}(y,x)G^{db}(x,y)\Bigr\}\bar{\phi}^{b}(y)
\label{IR2P8} 
\end{align}
where the trace is over the internal space
indices.  In the above expression the first term represents the
graph where two loops are joined together in a `figure eight' and
the second term represents the `setting sun' graph.  Using the
projection operators defined earlier this expression takes the
simplified form 
\begin{align}
	\G_{\textsc{e}}^{(2)}[\bar{\phi}]&={\l \over 4!}\int
d^{D}x \sqrt{g}\;\Bigl[3G^{2}_{1}(x,x)+
2(N-1)G_{1}(x,x)G_{2}(x,x)+  (N^{2}-1)G^{2}_{2}(x,x)\Bigr] \nn\\
&\qquad\qquad- {\l^{2} \over 36}\int d^{D}x \sqrt{g}\,\int d^{D}y
\sqrt{g}\,\bar{\phi}(x)\Bigl[3G_{1}^3(x,y)+(N-1)G_{2}^3(x,y)\Bigr]\bar{\phi}(y) \,.\label{IR2P9}
\end{align} 
It is evident from this expression that the dominant
contribution in the large $N$ limit comes from the `figure eight'
graph and from the $G_{2}$ propagator.  The $G_{2}$ propagator also gives the dominant contribution to the one loop effective action in the large $N$ limit, as can be seen from (\ref{IRON7}).

We now  evaluate the two particle irreducible  effective
action at the two loop level in the large $N$ limit.  Recalling
the definition of $\tilde{\G}_{\textsc{e}}[\bar{\phi},G]$ and noticing that the
`figure-eight graph is two particle irreducible we have (we also
rename $G_{2}\rightarrow G$ which is now a dressed propagator to
be solved for variationally along with the background field) 
\begin{align}
	\G_{\textsc{e}}[\bar{\phi},G] &= S_{\textsc{e}}[\bar{\phi}]+{N\over 2}\mbox{Tr}\ln
G^{-1}+{N \over 2}
\int d^{D}x \sqrt{g}\,A_{2}G(x,x)-{N \over 2}\mbox{Tr}({\bf 1})+{\l  \over 4!}\int d^{D}x
\sqrt{g}\,G^{2}(x,x) \label{IR2P10} 
\end{align}
where we have kept only
the dominant $O(N^{2})$--term from the two loop contribution
(notice that this term is also proportional to $\l$).

Let us pause here and ask how the usual effective action $\G_{\textsc{e}}[\bar{\phi}]$ is related to this newly introduced  $\G_{\textsc{e}}[\bar{\phi},G]$.  The answer is, when the current $K$  vanishes. Equivalently, it is 
$\G_{\textsc{e}}^{(2)}[\bar{\phi},G]$ at that value of $G(x,y)$ for which we
have \be {\d \G_{\textsc{e}}[\bar{\phi},G_{0}] \over \d
G(x,y)}=0\qquad\qquad\Rightarrow\qquad\qquad
\G_{\textsc{e}}[\bar{\phi}]=\G_{\textsc{e}}[\bar{\phi},G_{0}] \label{IR2P11} \te 
The difference is, $\G_{\textsc{e}}[\bar{\phi},G]$ contains  only two particle irreducible graphs, that is, graphs that do not split into disconnected nontrivial graphs upon opening two lines, while  $\G_{\textsc{e}}[\bar{\phi}]$ contains one particle
irreducible graphs.  Since $\G_{\textsc{e}}[\bar{\phi},G]$ does not contain
all the graphs that are included in $\G_{\textsc{e}}[\bar{\phi}]$, the
remaining ones must be accounted by the fact that the lines in a
graph belonging to $\G_{\textsc{e}}[\bar{\phi},G]$ represent the dressed
propagator $G$ and not the undressed one.  From this line of
reasoning, all the daisy graphs (they are two particle reducible,
and dominate the large $N$ contribution) must be contained in the
dominant two loop graph of $\G_{\textsc{e}}[\bar{\phi},G]$ since any daisy
graph can be built from the same graph, with the lines
representing the undressed propagator, by partially dressing the
propagator lines.

From the variational equation (\ref{IR2P11}) using our approximate effective action
(\ref{IR2P10}) we find (dropping the subscript from $G_{0}$) \be
G^{-1}(x,y)=-A_{2}(x,y)-{\l N \over 6}G(x,x)\d(x,y). \label{IR2P12}
\te Since \be A_{2}(x,y)=\left(-\Box_{\textsc{e}}+{M_{2}}^{2}\right)\d(x,y)
\label{IR2P13} \te 
[$M_2$ was defined in  \eqref{M2}.]  If we write 
\be G^{-1}(x,y)=\left(\Box_{\textsc{e}}-\chi\right)\d(x,y) \label{IR2P14} \te then
we have the consistency equation for the single component (no
internal indices) scalar field $\chi$: \be
\chi(x)=M_{2}^{2}+{\l N \over 6}G(x,x). \label{IR2P15} \te We
can now use (\ref{IR2P12}) to eliminate $A_{2}$ from
(\ref{IR2P10}) to obtain \be \G_{\textsc{e}}[\bar{\phi},G]=S_{\textsc{e}}[\bar{\phi}]+{N
\over 2}\mbox{Tr}\ln G^{-1}-{\l N^{2} \over 4!} \int d^{D}x
\sqrt{g}\;G^{2}(x,x) \label{IR2P16} \te Using (\ref{IR2P14}) and
(\ref{IR2P15}) with $\chi$ constant,  this becomes \bea
\G_{\textsc{e}}[\bar{\phi},\chi] &=& S_{\textsc{e}}[\bar{\phi}]+{N \over 2}\mbox{Tr}\ln\Bigl[(-\Box_{\textsc{e}}+\chi)\mu^{-2}\Bigr]-{3 \over 2\l}\int d^{D}x
\sqrt{g}\,\left(\chi-M_{2}^{2}\right)^{2} \label{IR2P17} \tea
The position-dependent effective mass $\chi$ satisfies the
iterative equation \be \chi={M_{2}}^{2}+{\l N \over
{6(\mathrm{Vol})}}\mbox{Tr}(-\Box_{\textsc{e}}+\chi)^{-1}, \label{IR2P18} \te 
where   $\mathrm{Vol}$  is the spacetime volume. This can
in principle be solved for $\chi=\chi(\bar{\phi})$ and when
substituted back into (\ref{IR2P17}) gives the one particle
irreducible effective action $\G_{\textsc{e}}[\bar{\phi}]$ with the
contribution of all daisy type graphs included. When $\bar\phi$
and $\chi$ are constants, (\ref{IR2P17}) defines a 2-particle
irreducible effective potential
\begin{equation}
V^{2PI}_{\rm eff}(\bar{\phi},\chi) = V(\bar{\phi})-{3 \over
2\l}(\chi-{M_{2}}^{2})^{2} +{N \over
2(\mathrm{Vol})}\mbox{Tr}\ln\Bigl[(-\Box_{\textsc{e}}+\chi)\mu^{-2}\Bigr].
 \label{IR2P19}
\end{equation}


Having derived the equations for the effective mass $M_{\rm eff} = \chi$ (\ref{IR2P18}) and the effective potential $V_{\rm eff}$ (\ref{IR2P19}) 
we now  study the symmetry behavior by seeking solutions to them for different geometries of interest. This will be followed by some remarks on the nature of finite-size effect in curved space related to its counterpart in condensed-matter physics, which characterizes the nature of IR in dS.

\subsection{Symmetry behavior in product spaces}
\label{sub-IRPS}

For spacetimes with topology $R^c \times B^b$ (where $B$ is a
$b$-dimensional compact space) $\chi$ is simply
\begin{equation}
\chi =M^2_2 + \frac{\lambda N}{6\Omega (B)} \int
\frac{d^c  k}{(2\pi)^c} \sum_{n} \frac{1}{k^2+\kappa_n + \chi} \;,
\label{IRPS1}
\end{equation}
where $\Omega (B)$ is the volume of the subspace B and $\kappa_n$
are the eigenvalues of $\Box_{\textsc{e}}$ restricted to $B$.  Considering only
the dominant lowest mode contribution $ (\kappa_0=0)$, we obtain
from integrating this lowest band the general expression for the
effective mass:
\begin{equation}
\chi =M^2_2 + \frac{\lambda N}{6\Omega(B) \chi} \left(\frac{\chi}{4\pi}
\right)^{c/2} \Gamma (1-c/2) \;
\label{IRPS2}
\end{equation}
which, expressed explicitly for each individual dimension, reads:
\begin{align}
	\chi &= M^2_2 + \frac{\lambda N}{6\Omega(B)\chi}  &&\text{for} &c&=0,
\label{IRPS3}\\
	\chi &= M^2_2 + \frac{\lambda N}{12\Omega (B)} \chi^{-1/2}  &&\text{for} &c&=1,
\label{IRPS4}\\
	\chi& = M^2_2 - \frac{\lambda N}{24\pi \Omega (B)} \left[\ln
\left(\frac{\chi \mu^{-2}}{4\pi} \right) + \gamma _E \right] &&\text{for} &c&=2,
\label{IRPS5}\\
	\chi &= M^2_2 - \frac{\lambda N}{24 \pi \Omega (B)} \chi^{1/2}  &&\text{for} &c&=3.
\label{IRPS6}
\end{align}
From these formulas we see that the solution $\chi=0$ for $M^2_2=0$
is not permissible for $c\leq 2$.  This can be interpreted to mean
that there cannot be a second-order phase transition for an $O(N)$
model in two or less dimensions (or, as Coleman  
put it: ``There are no Goldstone bosons in two or less dimensions.''), contained in the so-called Coleman-Mermin-Wagner theorem.  

Let us consider each case individually.

For $c=0$ we are dealing with a compact space with finite volume, e.g., $S^4$,  the Euclidean de Sitter, the case of special interest in this chapter.  The effective mass or
inverse correlation length is  obtained by solving (\ref{IRPS3}) when
$M^2_2=0$:
\begin{equation}
M^2_{\rm eff, dyn} = \chi =\left(\frac{\lambda N}{6\Omega}\right)^{1/2}
\label{IRPS7}
\end{equation}
for $N$-component scalar fields to leading order in large $N$. For $S^4$ the Euclidean `spacetime' volume $\Omega (S^4) =\mathrm{Vol} = 8\pi^2a^4/3$. This expression was first obtained by Hu and O'Connor \cite{HuOC87} -- the  effective mass  is  considered as dynamically-generated, thus also  referred to by later authors as  $M_{dyn}$. 
It  depends on the size of the system and vanishes as $\Omega \rightarrow
 \infty$.    A critical point which
should exist in the infinite-volume system will disappear in a
finite-volume theory: finite-size effect thus precludes a
second-order phase transition from occurring in finite systems.
This can be seen also from the form of the effective potential.
For massless, minimally coupled scalar fields in $S^4$ near
$\bar{\phi}=0$, the effective potential
obtained by inserting the solution of (\ref{IRPS3}) for $\chi$
with $M^2_2=(\lambda /6)\bar{\phi}^2$ into (\ref{IR2P19}) is
given by
\begin{equation}
V(\bar{\phi})=\frac{8\pi ^2}{3\Omega} \alpha(\lambda)_{IR} +
\frac{1}{2} \left(\frac{\hbar \lambda N}{6\Omega}\right)^{1/2} \bar{\phi}^2 +
\frac{\lambda}{48} \bar{\phi}^4 + \ldots \; \; , \label{IRPS8}
\end{equation}
where
$$
\alpha_{IR}(\lambda)=-\frac{3\hbar N}{32 \pi^2}\Bigl[1+\ln\frac{6\Omega a^4}{\hbar \lambda
N}\Bigr] \, .$$ 
Note that the $\ln \lambda$ dependence in \eqref{IRPS8}  
of $\alpha_{IR}$ renders the effective potential $V_{\bar\phi}$ singular  in the limit
$\lambda \rightarrow 0$.  This means that there does
not exist a ground state for the free ($\lambda=0)$ theory, but
such a state does exist for the self- interacting field.  Thus the
free massless minimally- coupled field is not an appropriate starting point
for perturbation theory.  One can, in fact, make the more general
statement that the same effect occurs for any field theory
defined in finite 4-volume for which the lowest eigenvalue of the
fluctuation operator is zero.  In the interacting case the
effective potential has to have positive curvature near
$\bar{\phi}=0$ and the minimum value of this effective mass
(inverse correlation length) is \eqref{IRPS7}. 
The power $1/2$ dependence is characteristic of a finite 4-volume situation
as is $2/3$ for a configuration with one-infinite dimension $(R^1
\times S^3)$.

From the above analysis it is apparent that there is always a local
minimum of the effective potential at $\bar{\phi} =0$ for the
massless minimally coupled field.  The  implication of this in critical phenomena is that  there cannot be a second order phase transition in  a finite volume.

As the volume becomes large in comparison with the volume sustained by the
correlation length, the effect of the finite size of the system
decreases and the higher mode contributions to the effective
potential can no longer be neglected.  To extract this large
volume behavior a different approximation which treats the system
as a small deviation from the infinite-volume limit is
necessary.   In general the approach to the infinite-volume
limit depends on the higher modes.  For example, in the case of
$S^4$, in the large-$a$ limit, the deviations due to the finite
size of the 4-sphere $S^4$ drop off as inverse powers of $a$; by contrast, for the 4-torus
$T^4$ the finite-size effect drops off exponentially.  The
effective potential for $T^4$ in the large volume limit is given by  
\begin{equation}
V_{\rm eff}(\bar{\phi}) = V_{\rm eff}^\infty (\bar{\phi}) - \frac{N
\chi^2}{2(2\pi^3)^{1/2}}\left[(\chi L_1^2)^{-5/4}e^{-(\chi
L_1)^{1/2}} + \ldots + (\chi L_4^2) ^{-5/4} e^{-(\chi
L_4)^{1/2}}\right] + \ldots \; \; . \label{IRPS9}
\end{equation}
The exponential fall-off seems to be characteristic of periodic boundary conditions (i.e., of the torus) and not a generic property of the approach to the bulk system;
it depends on the boundary conditions in the finite-volume setting 
\cite{BreZin85}

Considering also the other three categories $c=1, 2, 3$ (e.g., $c=1, b=3$ describes the Einstein universe, $c=3, b=1$   the imaginary-time finite temperature quantum field theory) Hu and O'Connor derived the forms of the 2PI  effective potentials  and the effective masses. See \cite{HuOC87} for the details .

The next question that naturally arises is what happens when the
higher modes are taken into account?  Can the theory develop a
minimum away from $\bar{\phi} = 0$ and thus allow the possibility
of a first order transition occurring in place of the second order
one?  Physically,  we know from condensed matter physics   that phase transitions which were second order in the bulk  can become first order in the finite volume setting.
Technically, to explore this possibility we need an expression for the potential that interpolates between the large volume limit and the result for small $\bar{\phi}$.   The leading behaviour of the effective potential is given by the effective potential of the lowest mode plus the loop contributions to the
effective potential of the higher (inhomogenous) modes, the latter contributions are treated in the usual loop expansion fashion.  When one goes beyond one loop corrections  there are, of course, interactions between the lowest mode and the higher modes.  It is considered in \cite{HuOC87} that the leading contributions from these interactions are taken into account by substituting the solution for $\chi(\bar{\phi})$ back into the higher mode contribution to the 2PI effective potential. 

How would these features  in Euclidean dS, namely, the IR dominance of the zero-mode, the contributions of the higher modes, manifest in  the Lorentzian description?   We will show some recent results of L\'opez Nacir et al \cite{LMT18} in  a later section for the  $O(N) \lambda \phi^4$ theory  in Lorentzian dS by means of a double expansion in $\sqrt \lambda$ and $1/N$ using functional methods.  

As for other fields, scalar electrodynamics (SQED) in de Sitter space  is believed to be a more realistic class of theories  for the description of phase transitions   in   inflationary cosmology.  O'Connor \cite{OCPhD} derived  the one-loop effective potential of scalar QED in de Sitter universe via the zeta function method.  His results 
were used to explore the infrared behavior of SQED and  draw implications about Planck scale phase transitions in \cite{HuOC86}. More details of this theory in dS space can be found in the work of Prokopec, Tsamis and  Woodard \cite{ProTsaWoo08}.  Functional renormalization group methods have been applied to SQED by Gonz\'alez  and Prokopec to obtain quantum loop effects \cite{GonPro16}. See also  the study of the  critical behavior of a $\phi^4$ theory in spherical and hyperbolic spaces by  Benedetti \cite{Ben15}.

\section{Lorentzian de Sitter: late time IR  and Stochastic Approach}





For de Sitter space in the $S^4$ coordinatization, we learn in the previous two sections how to analyze the infrared behavior of interacting quantum fields by way of the 2PI effective potential under the large $N$ approximation, and the physical meaning of such behavior, in terms of finite size effect and infrared dimensional reduction.  
Now we ask the question: If one views de Sitter spacetime in a dynamical setting, such as in the $k=0$ or $1$ RW coordinatizations, with topology $S^3 \times R^1$ or $R^3 \times R^1$, the compactness of $S^4$ is gone, would there still be a finite size effect? Would we expect different physics? Or, rather, since physics should be the same despite the differences in coordinate representations, how would the physics manifest in a dynamical setting? If we call the $S^4$ treatment as Euclidean, we will refer to the latter two cases as Lorentzian, since time evolution is explicit.

\subsection{Lorentzian-de Sitter:  Dynamical finite size effect}
The resolution of this puzzle brings in an interesting point related to the
effect of spacetime dynamics on the symmetry behavior of a quantum field.
Specifically, for the special case of exponential expansion,
in the spectrum of the 4-dim (spacetime) wave operator,
there is a gap which separates the zero mode from the rest. This is what
gives the effective infrared dimension  $d_{\rm eff}\simeq 0$ for the deep IR behavior of the scalar field in these other coordinate descriptions.
Physically this arises from the fact that, as a result of exponential expansion
the whole spectrum undergoes rapid red-shifting, and given long enough time (with only logarithmic difference) almost all the higher modes will heap upon the zero mode. (The $S^3$ or $R^3$ spatial sector also becomes immaterial.) That is why at late times the longest wavelength mode gives the dominant contribution to  the system's IR behavior.  The
appearance of a scale (the event horizon $H^{-1}$)
is a unique feature  of this exponential class of expansion.
It gives rise to effects similar to those originating from the existence of some
finite size in static spacetimes. 

\subsubsection{Inflation as Scaling: Eternal inflation as static critical phenomena}

One can see the linkage between the Euclidean- and the RW- de Sitter formulations  from the observation \cite{cgea} that the exponential
expansion of the scale factor can be viewed as a system undergoing a (Kadanoff-Migdal) scale transformation \cite{Wil83}.
In a  spatially-flat RW metric with a constant scale factor --this is just the Minkowski spacetime --  let us consider an ordered sequence of such static hyperspaces (foliation) with scales $a_{0}$, $a_{1}$, $a_{2}$, etc parameterized by $t_{n} = t_{0} + n \Delta t$, $n = 0, 1, 2, \ldots $ .
These spacetimes have the same geometry and topology but differ only in the physical scale in space.  One can always redefine the physical scale
length $x'_{(n)} = a_{n} x$ to render them equivalent.  If each copy has scale length magnified by a fixed factor $s$ over the previous one in
the sequence, i.e. $a_{n+1}/ a_{n} \equiv s =e^{H\Delta t}$, we get exactly the physical picture as in an eternal inflation. 
After $n$-iterations i.e.
$a_{n}/a_{0} = e^{n(H\Delta t)}$, or, with a continuous parameter
$a(t) = a_{0}e^{Ht}$.  It is
important to recognize that $t$ can be any real parameter not necessarily
related to time.  In other words, time in eternal inflation plays the role of
a scaling parameter.  Dynamics in this case is replaced by scaling, and  every time step of eternal inflation can be viewed as a scaling transformation in a static setting.

In this light we see that in  the `dynamic'  RW description of de Sitter, the constant time scale associated with exponential expansion, $1/H$, corresponds to the length scale in an equivalent static universe, the radius of $S^4$ in the Euclidean  de Sitter space.   In the Lorentzian QFT  the infrared behavior is dominated by the late time behavior of the longest wavelength mode, and that is equivalent to the infrared behavior of the zero-mode in $S^4$ in a `static' description.  The IR physics in Lorentzian dS is called dynamical finite size effect by Hu and O'Connor \cite{HuEdmonton}, as the dynamical correspondence of finite size effect in a compact space. 

As a reminder of the relation between the Euclidean de Sitter (EdS) and the Lorentzian de Sitter (LdS)  metrics   and to establish notations used in the sections below, we start with the global coordinate which covers the full  $d$-dim LdS spacetime,
\begin{equation}
 ds^2 =-  dt^2 + \frac{1}{H^2} \cosh^2(H t) d\Omega^2_{d-1}. \label{dLdS}
\end{equation}
Analytically continue $t$ to imaginary time $t \to i ( \tau  - \pi/2H )$  with the periodicity condition $\tau = \tau + 2\pi H^{-1}$   imposed  to avoid a multivalued metric. The resultant EdS is a $d$-sphere of radius $H^{-1}$
\begin{equation}
 ds^2 = H^{-2} \left[ d \theta^2 + \sin(\theta)^2 d\Omega^2 \right], \label{d-sphere}
\end{equation}
with $\theta = H\tau$.

We now turn to the  stochastic approach of Starobinsky and Yokoyama  \cite{StaYok94}.  We shall see at the end that the infrared behavior induced by  the  dynamical effects of exponential expansion at  late times -- one very clever way to express this in terms of stochastic fields was proposed  by   Starobinsky and Yokoyama \cite{StaYok94} ---  is equivalent to the finite size effect associated with the zero mode in the Euclidean formulation described by Hu and O'Connor \cite{HuOC86,HuOC87}. 


\subsection{Stochastic Approach for long-wavelength late-time behavior}

Working within the structural framework of stochastic inflation model proposed by Starobinsky ~\cite{Sta86} in 1986  [Note A].  
Starobinsky and Yokoyama \cite{StaYok94} in 1994  calculated the two-point function of the  long-wavelength field at late times.  This result captured the attention of quantum field theorists working on inflationary cosmology because it is a nonperturbative result derived from a very different method.  As we   explained earlier, this can be understood from the  dominance of the zero mode contribution to the infrared behavior of quantum fields in Euclidean dS.  Beneke \& Moch \cite{BenMoc13} gave a clear derivation of the late time two-point functions in the stochastic scheme and  showed the relation between these two approaches.   We summarize their  derivations below.  

Divide the scalar field into two sectors $\phi=\bar\phi+\phi_<$, where $\bar\phi$ contains all  long-wavelength modes with wave number $k < \epsilon a H$ ($a$ is the scale factor and $\epsilon \ll 1$ is a parameter which separates the long wavelength from the short wavelength modes), assumed to be superhorizon which behaves classically and a short wavelength subhorizon part  $ \phi_<$ which retains the quantum field character but whose effect can be represented by a classical noise, assumed to be white. [Note B]\\

[Note A]  Some cautionary notes on the structural foundation of stochastic inflation may be helpful here.  As pointed out in \cite{CalHu95}, two issues of fundamental concern are:   1) the decoherence of the long wavelength modes and 2) the justification of a white noise for the short wavelength sector.  For issue 1) the justification for treating the long wavelength  sector as  classical  need be provided.  Usually one invokes decoherence, namely, that the long wavelength modes are readily decohered by the short wavelength or high frequency ($>$) sector,  treated as noise via some environment-induced decoherence mechanism. The decoherence scheme of Polarsky and Starobinsky \cite{PolSta96} asserting that the insignficance of the decaying mode renders the growing mode effectively decohered is too simplistic,  because quantum wave functions even at   small attenuated amplitude has a chance to recohere.   A later version with Kiefer \cite{KiePolSta98} comes closer to the open-system treatment such as done in \cite{KMH97}. A first example of cosmological decoherence was given in \cite{HPZ93Dec} using an interacting field theory based on the insight gained from the study of  nonMarkovian processes in quantum Brownian motion models \cite{HPZ92}.   Lombardo and Mazzitelli \cite{LomMaz96}  presented an open system treatment of decoherence in  interacting quantum fields based on the use of the coarse-grained effective action introduced in \cite{cgea} and the influence functional formalism \cite{FeyVer63}.  For its relation to effective field theory, see \cite{BHTW15}.

[Note B]:  For the noise issue 2),  a sharp  partitioning of the field into two sectors does not constitute an interaction between the two sectors,  whereas an interaction is necessary for an environment (short wavelength sector) - induced decoherence (of the long wavelength sector) to occur. Winitzki and Vilenkin  \cite{WinVil00} examined the window function and expressed doubt that the sharp division generates  a white noise spectrum.   This is problematic from a quantum field theory viewpoint because in an expanding background the Hilbert space of both sectors are changing in time,  and a  proper treatment of their interaction  is rather involved.   The  open quantum systems treatment for noise and environment-induced decoherence provided the basis for later more vigorous investigations carried out by e.g., Lombardo and L\'opez Nacir  \cite{LomNac05}.  (For further description of this topic,  see, e.g.,  Chap 15 of \cite{CalHu08}.)\\

 Now suppose that $\phi$ is acted on by a scalar field potential $V(\phi)$. From the field equation for $\phi$, namely, $-\Box \phi + V '(\phi) =0 $ we obtain an equation governing $\bar\phi$ in the form of a Langevin equation 
\begin{equation}
\dot{\bar\phi}(t,\mathbf{x}) = -\frac{1}{(d-1) H} \,V^\prime(\bar\phi) + \xi(t,\mathbf{x}),
\label{dlangevin}
\end{equation}
where $\xi(t,\mathbf{x})$ denotes the stochastic force generated by the short-wavelength modes 
\begin{equation}
\xi(t,\mathbf{x})= \dot{\phi_<} = \epsilon a H^2 \int
\frac{d^{d-1} k}{(2\pi)^{d-1}}\,\delta(k-\epsilon a H)\, 
\left(a_k \phi_k(t) e^{i \mathbf{k}\cdot\mathbf{x}} + \mbox{c.c.} \right),
\end{equation}
where $\phi_k(t)$ are the mode functions.  
At leading order, we can neglect the self-interaction of the short-wavelength modes whence the fluctuations satisfy 
\be \langle \xi (t_1,\mathbf{x}) \xi (t_2,\mathbf{x})  \rangle =\gamma\; \delta(t_1-t_2) \ee with 
\begin{equation}
\gamma = \frac{2 \pi^{\frac{d-1}{2}}}{(2\pi)^{d-1} 
\Gamma(\frac{d-1}{2})} 
\times \frac{2^{d-3} \,[\Gamma(\frac{d-1}{2})]^2}{\pi}\times H^{d-1}.
\end{equation}
The first factor arises from the volume of the $d-2$ dimensional  momentum shell $k=\epsilon a H$, the second from the 
long-wavelength limit (since $k =  \epsilon a H \ll a H$) of the Bunch-Davies mode functions $\phi_k(t)$. 
The Fokker-Planck equation associated with (\ref{dlangevin}) for the one-particle probability density ${\cal P}[\varphi]$ has the form\footnote{In this context $\varphi$ is a stochastic variable in contrast to it denoting the fluctuations of a quantum field variable used in earlier sections.} 
\begin{equation}
\frac{\partial {\cal P}}{\partial t} = \frac{1}{(d-1) H} \,
\frac{\partial}{\partial\varphi}\,(V^\prime(\varphi){\cal P}) 
+\frac{\gamma}{2}\,\frac{\partial^2}{\partial\varphi^2}\,{\cal P}.
\label{fokkerplanck}
\end{equation}
At late times it admits a stationary (`equilibrium') solution 
\begin{equation}
{\cal P}[\varphi] = {\cal N}\exp\left(-\frac{2}{(d-1)\gamma H}\,V(\varphi)\right),
\end{equation}
where the normalisation ${\cal N}$ is determined by the condition  $\int\limits_{-\infty}^\infty d\varphi\; {\cal P}[\varphi]=1.$
In terms of this  the two-point function of the long-wavelength field at late times is given by
\begin{equation}
\langle \bar\phi\bar\phi \rangle = {\cal N}\int_{-\infty}^\infty\! d\varphi \;
\varphi^2  \exp\left(-\frac{2}{(d-1)\gamma H}\,V(\varphi)\right).
\end{equation}
As shown by Beneke and Moch \cite{BenMoc13},
this expression for the long-wavelength two-point functions (with $V(\phi) = \frac{\lambda}{4!}\phi^4$), and the `dynamical masses' derived from it, 
agrees with the exact Euclidean result at leading order in the expansion in  $\sqrt{\lambda}$  
provided the dissipation and fluctuation coefficients in the Fokker-Planck equation are related to the volume of $d$-dimensional Euclidean de Sitter space with radius $1/H$ by $(d-1) H \times\frac{\gamma}{2} = \frac{1}{V_d}$, which can be easily verified. 




Furthering this,   Garbrecht, Rigopoulos, and  Zhu  \cite{GarRigZhu14} found a perturbative diagrammatic representation of the stochastic function in the Starobinsky-Yokoyama approach which is worthy of mentioning, as follows: 

In the stochastic approach, the long wavelength modes $\bar \phi$ of the scalar inflaton field $\phi$ obey a Langevin-type equation 
\be\label{Langevin}
\dot{\bar\phi}+\frac{ V'(\bar\phi)}{3H} = \xi(t)\,,
\ee
where $\xi$ is a Gaussian random force with
\be \label{noise}
\langle\xi(t)\xi(t')\rangle = \frac{H^3}{4\pi^2}\delta(t-t')\,.
\ee
At late times the stochastic process (\ref{Langevin}) is described by the probability distribution function~\cite{Sta86,StaYok94}
\be
\label{stoch:PDF}
{\cal P}[\bar\phi]={\cal N}\;{ e}^{-\frac{8\pi^2}{3H^4} V(\bar\phi)}\,.
\ee
Following Starobinsky's original approach, expectation values (at equal times) are obtained by weighing the variable by the probability distribution function, e.g.,
\be
\label{phi:expectationvalue}
\langle \bar\phi^n \rangle
=\int\limits_{-\infty}^\infty d\bar\phi\; \bar\phi^n {\cal P} [\bar\phi]\,.
\ee
For a potential of the form
\be
\label{potential}
V(\phi)=\frac 12 m^2 \phi^2 +\frac{\lambda}{4!}\phi^4\,.
\ee
one can carry out the expansion
\be
{\rm e}^{-\frac{8\pi^2}{3H^4}V(\phi)}
={\rm e}^{-\frac{4\pi^2}{3H^4} m^2 \phi^2}
\left(
1-\frac{\pi^2 \lambda \phi^4}{9H^4}+\frac{\pi^4\lambda^2\phi^8}{162 H^8}+\cdots
\right)\,.
\ee
Substituting this into Eq.~(\ref{phi:expectationvalue}), performing the integrals for the fluctuation and the normalization leads to the result
\be
\label{fluc:massive}
\langle \phi^2 \rangle=\frac{3H^4}{8\pi^2 m^2}-\frac{9\lambda H^8}{128\pi^4m^6}+\frac{9\lambda^2 H^{12}}{256 \pi^6 m^{10}}+\cdots\,.
\ee

To bring the stochastic approach in a form closer to what can be checked against QFT methods, GRZ \cite{GarRigZhu14} introduced a functional representation of  Eq (\ref{Langevin}) and (\ref{noise})  from which the expectation values of an operator  can be calculated  by means of distribution averages. In this framework, the expectation value of an operator $\mathcal{O}[\phi]$ is given by \footnote{The functional integration measures that appear in Eq. (\ref{funcexpstoch}) is such that $<\mathbf{1}>=1$. This corresponds to the retarded Ito regularization of the stochastic equation (\ref{Langevin}) \cite{KamLev09}.}
\beq
\langle {\cal O}[\phi] \rangle = \int\! D\!\left[\xi\right]\; { e}^{-\frac{1}{2}\int \!\! dt\,\xi^2\frac{4\pi^2}{H^3}}\int\! D\!\left[\phi\right]\;{\cal O}[\phi]\,\delta\!\left(\dot{\phi}+{\partial_\phi V}/{3H} - \xi\right)\,.\label{funcexpstoch}
\eeq
By expressing the delta functional as a functional ``Fourier transform'' with the aid of an auxiliary field $\psi$,  setting a functional determinant that appears in the integration to unity and performing the Gaussian $\xi$ integral GRZ obtain
\beq\label{GenFunc}
\langle {\cal O}[\phi] \rangle=\int\! D\!\left[\phi, \psi\right]\; {\cal O}[\phi] \,{\rm e}^{-\int \!\!dt \,\left[ {\rm i}\psi\left(\dot{\phi}+\frac{ \partial_\phi V}{3H}\right) + \frac{H^3}{8\pi^2}\psi^2\right]}\,.
\eeq
To obtain a diagrammatic expansion one can rewrite the action by bringing the quadratic part in a more symmetric form.  In fact, the analogue of Eq.~(\ref{GenFunc}) containing second time derivatives can be obtained directly from the more fundamental closed-time-path (CTP) integral after short wavelength modes are integrated out~\cite{GGRZ15}.

To construct the two point function of $\phi$, we begin with the retarded and advanced Green functions $G^{(R,A)}(t,t')$  for the operator $\partial_t +\frac{m^2}{3H}$,
\beq
G^R(t,t')=G^A(t',t)=
{\rm e}^{-\frac{m^2}{3H}(t-t')}\Theta(t-t')\,.
\eeq
The 2-point function $F(t,t')$ of $\phi$  is then  given by
\bea
F(t,t')\equiv\langle\phi(t)\phi(t')\rangle &=& \frac{H^3}{4\pi^2}\int\limits_{0}^{+\infty} d\tau \,\,G^R(t,\tau)G^A(\tau,t') \nn\\
&=&\frac{3H^4}{8\pi^2m^2}\left({\rm e}^{-\frac{m^2}{3H}|t-t'|}-{\rm e}^{-\frac{m^2}{3H}(t+t')}\right)\,.\label{corr}
\eea
If $t$ and $t'$ are taken to be sufficiently large, or, equivalently, the stochastic process is taken to have begun early enough, the correlator reduces to
\beq\label{corr2}
F(t,t')\simeq \frac{3H^4}{8\pi^2m^2}{\rm e}^{-\frac{m^2}{3H}|t-t'|}.
\eeq
Note that in the massless limit, $G^R(t,t')\rightarrow \Theta(t-t')$, and the variance, already known in \cite{Vil83NPB}, grows linearly with time
\beq
\langle\phi^2(t)\rangle_{m=0}\simeq \frac{H^3}{4\pi^2}t.
\eeq

A diagrammatic representation of the stochastic processes can be found in   Fig. 3 of \cite{GarRigZhu14}.  At  late times  the equal time two-point function  $\lim\limits_{t\rightarrow\infty}\langle\phi(t)^2\rangle$  has the same form as (\ref{fluc:massive}) obtained via the probability functional in the stochastic method.  (Indeed the three terms in  (\ref{fluc:massive})  can be obtained from Feynman diagrams by the following rules: For each vertex, assign a factor  $-\lambda 8\pi^2/(3H^4)$ and for each propagator, a factor $3H^4/(8\pi^2 m^2)$, then divide by the appropriate symmetry factor. Compare this with the field-theoretical diagrams in Fig 1 of  \cite{GarRigZhu14} )   

 GRZ  showed that the perturbative results obtained from perturbative calculations for small massive scalar field 
agrees with the result from the QFT Schwinger-Dyson equations. Each individual contribution from the topologically distinct diagrams  
equals the QFT contribution from diagrams of corresponding topology.
GRZ  have also considered a massive theory with the mass of the scalar field $\phi$ chosen to be small enough, such that the amplification of superhorizon momentum modes leads to a significant enhancement of infrared correlations, typical of the massless case, but large enough such that perturbation theory remains valid. Using the closed-time-path  approach, they calculated the infrared corrections to the two-point function of $\phi$ to 2-loop order. To this approximation, they find agreement with the correlation found using the stochastic method. 

 Continuing the  investigations by GRZ for a light massive scalar field with $\lambda \phi^4$ self-interaction,  Garbrecht,  Gautier,  Rigopoulos, and  Zhu (GGRZ) \cite{GGRZ15} carried out a diagrammatic expansion that describes the field as driven by stochastic noise. This  is compared with the Feynman diagrams in the Keldysh basis of the closed-time-path field-theoretical formalism. For all orders in the expansion,  they find that the diagrams agree when evaluated in the leading  infrared approximation, i.e. to leading order in $m^2/H^2$, where $m$ is the mass of the scalar field and $H$ is the Hubble rate. As a consequence, the correlation functions computed in both approaches also agree to leading infrared order. This perturbative correspondence shows that the stochastic theory is exactly equivalent to the field theory in the infrared. The former can then offer a nonperturbative resummation of the field theoretical Feynman diagram expansion, including fields with $0 \leq m^2 \ll \sqrt \l  H^2$ for which the perturbation expansion fails at late times.

\subsubsection{Connecting the stochastic approach to  effective action method}

The  2PI effective action we introduced in Sec. 2  is in an Euclidean formulation. For  dS QFT an Euclidean formulation can be shown to be equivalent to the in-in formulation  \cite{HigMarMor11a}.  We shall present  in the next chapter the CTP (`in-in') 2PI  effective action  in Lorentzian QFT together with loop and $1/N$ expansion methods.  For the $O(N) \l \phi^4 $ quantum field theory  in a general curved spacetime this has been derived by Ramsey and Hu  at two loops, see 
Eqs (5.17) \& (5.23) in \cite{RamHu97ON}.  
From there they obtained a set of local, covariant, nonperturbative equations for the mean-field $\bar \phi$ and for the two-point function $C^{++}$ 
[Eq. (5.25) 
and  Eq. (5.24) respectively in \cite{RamHu97ON}]. 
At leading order $1/N$ expansion these equations become  
Eqs (5.34) and (5.32) in \cite{RamHu97ON}.     
Riotto and Sloth \cite{RioSlo08} used these prior results to investigate the infrared behavior of this field theory in  the flat-RW representation of the de Sitter universe. As they are interested only in the IR properties in these equation  they omitted  the appropriate counter terms needed to cancel the UV divergences.  The way they  solved the gap equation for long wavelengths is to first evaluate $C^{++}(x,x)$ from S\&Y's Fokker Planck equation, which is a good approximation for the late time behavior. They then inserted this into the gap equation and solved for $C^{++}(x,x')$,  obtaining a simple analytic form for it.  We will give a summary description of it in the next chapter as a sample application of these advanced techniques.  

While  Riotto and Sloth borrowed some results from the stochastic approach using the Fokker-Planck equation of S\&Y one can also adhere to the field theory route to calculate the higher order correlation function via a diagrammatic expansion of the in-in effective action. Petri \cite{Pet08} obtained a resuumation of the infrared divergences reproducing the results in \cite{RioSlo08}.  Treatment of the infrared problem via diagrammatic expansion of the partition function in the stochastic approach was further pursued in \cite{GarRigZhu14,GGRZ15}.
Finally we mention the work of Moss and Rigopoulos \cite{MosRig17}  
who, using the Schwinger-Keldysh path integral formalism, derived an effective potential which at leading order and  for time scales  $\D t \gg H^{-1}$  produces Starobinsky's stochastic evolution,  but which also allows for the computation of quantum UV corrections.  The long wavelength stochastic dynamical equations are now second order in time, incorporating temporal scales $\D t \sim H^{-1}$  and resulting in a Kramers equation for the probability distribution,  the Wigner function, in contrast to the more usual Fokker- Planck equation. This feature allows for a non-perturbative evaluation, within the stochastic formalism, not only expectation values of field correlators, but also the stress-energy tensor of the inflaton field $\phi$.

\subsection{Resummation of secular terms and long distance behaviour of $O(N)$ correlators in LdS}

The three main approaches to the IR problem of QFT in dS space are, as we have learned  in the above, the Euclidean, the Lorentzian and the stochastic formulations.  We have seen the development in Euclidean formulation in the second section and the stochastic formulation in this section, where the relation between the two are discussed.  Now let us return to QFT and see the relation between the Euclidean and the Lorentzian formulations in the depiction of IR in dS.   Linking these two approaches involves taking the zero mode in the Euclidean dS as providing the dominant contribution, but also including the nonzero modes in the Lorentzian formulation. We give a summary of  the recent work of L\'opez Nacir,  Mazzitelli and Trombetta (LMT) \cite{LMT18} who perform a resummation in a double expansion

In Euclidean dS,  recognizing that the dominant IR behavior comes from the zero-mode in the fluctuation operator of a quantum field on $S^4$ is an important first step \cite{HuOC86,Raj10,BenMoc13}. The effective or dynamically-induced mass  given by  \eqref{IRPS7}  gives to leading order in  $\sqrt\lambda$  the correlator of the IR part of the field,
  \begin{equation}\langle\phi_{IR} (x)\phi_{IR}(x')\rangle\simeq\frac{3 H^3}{8\pi^2 m_{\rm dyn}^{2}}\,.\end{equation}
The above nonperturbative result  cures the divergence appearing in the  free massless correlator. However,  corrections beyond  this constant  IR contribution  are necessary to  understand the long distance  limit  of  the  correlators in the Lorentzian dS (LdS) spacetime. Consider again the  $O(N)$-model with quartic self-interaction in an Euclidean dS space represented by a $d$ dimensional sphere of radius $H^{-1}$ around the  unbroken symmetric vacuum
\begin{equation}
 S_{\textsc{e}} =-i \int d^d x \sqrt{g} \left[ \frac{1}{2} \phi_a \left( -\square_{\textsc{e}} + m^2 \right) \phi_a + \frac{\lambda}{8N} ( \phi_a \phi_a )^2 \right], \label{Eaction}
\end{equation}
where $\phi_a$ are the components of an element of the adjoint representation of the $O(N)$ group, with $a=1$, $...$, $N$. The sum over repeated indices is implied\footnote{Beware of possible differences in notations, such as the factor multiplying the $(\phi_a\phi_a)^2$ term in \eqref{Eaction}.}.  
 In   \cite{LMT16}  a  systematic nonperturbative resummation scheme  has been developed for a double expansion in $1/N$ and $\sqrt\lambda$  for sufficiently large number of fields $N$.  Formulated in  EdS it consists of  a reorganization of the perturbative expansion followed by an analytic continuation to the LdS. The main advantages are:  the  improvement of the IR behavior of correlation functions in LdS is systematic,  and  the renormalization process is well understood. We summarize   the  simpler derivation    in    \cite{LMT18} below
 and discuss  the results  for the  long wavelength  limit  of the correlators up to next-to-leading  order (NTLO) in  $1/N$.

\subsubsection{Reorganizing the perturbative expansion}\label{SimplerRess}

LMT begin by separating the field into  into two sectors $\phi_a(x) =  \phi_{0a} + \tilde{\phi}_a(x)$,  the homogeneous zero mode  $\phi_{0a}$ and  the inhomogeneous modes  $\tilde{\phi}_a(x)$.  The former is treated nonperturbatively,  following  the method used in  ~\cite{HuOC87, Raj10,BenMoc13} for QFT in EdS.
 The interaction part of the action Eq.~(\ref{Eaction}) is separated accordingly, as follows:
\begin{equation}
 S_{int} = \frac{\lambda V_d}{8N}   |\vec{\phi}_0|^4 + S^{(2)}_{int}[\phi_{0a},\tilde{\phi}_a], \label{Sint2}
\end{equation}
where 
\begin{equation}
V_d =  \int d^dx \sqrt{g} = \frac{2\pi^{\frac{d+1}{2}}}{\Gamma\left(\frac{d+1}{2}\right)H^d}
\end{equation} is the total volume of EdS space equal to the surface area of the $d$-sphere, and  $S^{(2)}_{int}$ contains at least two powers of $\tilde{\phi}_a$ (note that the term linear in $\tilde{\phi}_a$ vanishes identically by orthogonality). 
The path integral over the constant zero modes is an ordinary integral, which can be performed exactly (i.e. nonperturbatively in the coupling constant $\lambda$). The generating functional becomes
\begin{align}
   Z[J_0, \tilde{J}] &= \mathcal{N} \int d^N\phi_0\! \int \mathcal{D}\tilde{\phi} \, e^{-S_{\textsc{e}} - \int_x  (\vec{J}_{0} \cdot \vec{\phi}_{0} + \tilde{J}_a \tilde{\phi}_a )} = \exp\left\{ -S^{(2)}_{int}\left[\frac{\delta}{\delta J_0},\frac{\delta}{\delta \tilde{J}}\right] \right\} Z_0[J_0] \tilde{Z}_{f}[\tilde{J}],
   \label{Z-int}
\end{align}
where $J_{0a}$ and $\tilde{J}_a$ are external sources and the shorthand notation $\int_x = \int d^d x \sqrt{g}$ is introduced. The zero part $Z_0[J_0]$ is defined as the exact generating functional of the theory with the zero modes alone  \cite{Raj10},
\begin{equation}
 Z_0[J_0] = \frac{\displaystyle\int\!d^N\phi_0 \; e^{-V_d \left[\frac{\lambda}{8N}  |\vec{\phi}_0|^4 + \frac{m^2}{2}  |\vec{\phi}_0|^2 + \vec{J}_{0} \cdot \vec{\phi}_{0} \right]}}{\displaystyle\int\! d^N\phi_0 \; e^{-V_d \left[ \frac{\lambda}{8N}  |\vec{\phi}_0|^4 + \frac{m^2}{2}  |\vec{\phi}_0|^2 \right]}}.
\end{equation}
For a massless field $m=0$, the variance of the zero modes is 
\begin{equation}
 \langle \phi_{0a} \phi_{0b} \rangle_0 = \frac{\displaystyle\int\! d^N\phi_0 \; \phi_{0a} \phi_{0b} \, e^{- \frac{V_d \lambda}{8N}  |\vec{\phi}_0|^4 }}{\displaystyle\int d^N\phi_0 \; e^{- \frac{V_d \lambda}{8N}  |\vec{\phi}_0|^4 }} = \delta_{ab} \sqrt{\frac{2}{V_d \lambda}} \frac{2}{\sqrt{N}} \frac{\Gamma\left( \frac{N+2}{4} \right)}{\Gamma\left( \frac{N}{4} \right)} \equiv \frac{ \delta_{ab}}{V_d m_{\rm dyn}^2},
\end{equation}
which allows the identification of a dynamically-generated effective mass $m_{\rm dyn}^2$  \eqref{IRPS7} \cite{HuOC87}.
 
This result is valid at LO in $\sqrt{\lambda}$ and for all $N$. Corrections coming from the inhomogeneous modes can then be computed by treating $S^{(2)}_{int}$ in Eq.~(\ref{Z-int}) perturbatively. The free  correlators  of the inhomogeneous modes  can be written as $ \tilde{G}^{(m)}_{ab}(r) =  \delta_{ab}    \tilde{G}^{(m)}$, with
\begin{equation}
 \tilde{G}^{(m)}(r) =    {G}^{(m)}(r) -\frac{1}{V_d m^2} =H^d \sum_{\vec{L},|\vec{L}|>0 } \frac{Y_{\vec{L}}(x) Y^*_{\vec{L}}(x')}{H^2 L(L+d-1) + m^2}, \label{free-prop} 
\end{equation} where  ${G}^{(m)}(r)$ is the standard  propagator for a free field with mass $m$ and $Y_{\vec{L}}(x)$ are the characteristic functions of the Euclidean space, e.g., the hyperspherical harmonics with $L=(n, l, m)$ in the case of $S^4$.  
Although the correlator    $\tilde{G}^{(m)}$ is finite  for $m \to 0$,
 the perturbation theory built with the massless $\tilde{G}^{(0)}$ is still ill-defined when analytically continued to LdS and at long distances/late times, due to the divergent behavior  given by  
\begin{equation}
 G^{(m)}(r) \simeq \frac{1}{V_d m^2}\;r^{-\frac{m^2}{d-1}}.
  \label{late-times-massive}
\end{equation}
 Solving this issue requires further resummations of contributions that also involve the inhomogeneous modes.
 A subclass of such contributions come from the  terms in $S^{(2)}_{int}$ that are quadratic in both $\tilde{\phi}_a$ and $\phi_{0a}$, which dress $\tilde{G}$ with a nonperturbative mass. 
 
\subsubsection{Resummation of bi-quadratic terms}
In the spirit of the separation of the interaction part of the action done in Eq.~(\ref{Sint2}), we further isolate the bi-quadratic terms,
\begin{equation}
 S^{(2)}_{int} = \frac{\lambda}{8N} \int d^d x \sqrt{g} \left[ 2 |\vec{\phi}_0|^2 |\vec \tilde{\phi}|^2 + 4 (\vec{\phi}_{0} \cdot \vec{\tilde{\phi}})^2 \right] + S^{(3)}_{int}, \label{Sint3}
\end{equation}
where now $S^{(3)}_{int}$ contains terms with at least three powers of $\tilde{\phi}_a$.
The main idea is  to include the bi-quadratic terms in the definition of the propagator $\tilde{G}$ by  defining
$\vec{\phi}_{0}$-dependent inverse propagators
\begin{equation}
 \tilde{G}^{-1}_{ab}(\vec{\phi}_0) = -\delta_{ab} \square + m_{ab}^2(\vec{\phi}_0)\,,
\end{equation}
with the following mass matrix
\begin{equation}
 m_{ab}^2(\vec{\phi}_{0}) = m^2 \delta_{ab} + \frac{\lambda}{2N} \left( \delta_{ab} \delta_{cd} + \delta_{ac} \delta_{bd} + \delta_{ad} \delta_{bc} \right) \phi_{0c} \phi_{0d} = m_1^2 \, P_{ab} + m_2^2 \, \epsilon_a \epsilon_b,
\end{equation}
where in the second line  the matrix  is split into the parallel and transverse components with respect to the $\epsilon_a \equiv \phi_{0a}/|\vec{\phi}_0|$ direction, by means of the projector $P_{ab} = \delta_{ab} - \epsilon_a \epsilon_b$. Here $m_1^2 = m^2 + \frac{\lambda}{2N} |\vec{\phi}_0|^2$ and  $ m_2^2 = m^2 + \frac{3\lambda}{2N} |\vec{\phi}_0|^2. $
 Diagonalizing the mass matrix $m_{ab}^2$, the generating functional becomes\begin{equation}
 Z[\vec{J}_0, \tilde{J}_i^{(1)}, \tilde{J}^{(2)}] = \exp\left(-S^{(3)}_{int}\left[\frac{\delta}{\delta J_0},\frac{\delta}{\delta \tilde{J}}\right]\right) \mathcal{Z}[\vec{J}_0, \tilde{J}_i^{(1)}, \tilde{J}^{(2)}],
\end{equation}
 where  $ \mathcal{Z} $ is a new ``free'' generating functional, 
\begin{eqnarray}
\mathcal{Z}[\vec{J}_0, \tilde{J}^{(1)}_i, \tilde{J}^{(2)}] &=& \mathcal{N} \Biggl\langle \sqrt{ \det \tilde{G}_1}^{N-1} \sqrt{ \det \tilde{G}_2} \, \tilde{Z}_1[\tilde{J}^{(1)}_i] \,  \tilde{Z}_2[\tilde{J}^{(2)}] \Biggr\rangle_{0}^{\vec{J}_0}  \nn\\
 &\equiv& \Bigl\langle \tilde{Z}_1[\tilde{J}^{(1)}_i] \,  \tilde{Z}_2[\tilde{J}^{(2)}] \Bigr\rangle_{\bar{0}}^{\vec{J}_0}.
\label{Z-LO}
\end{eqnarray}
Here we use the shorthand $\tilde{G}_\alpha \equiv \tilde{G}^{(m_\alpha)}$,  $\tilde{Z}_{\alpha} \equiv \tilde{Z}_{f}\left[\tilde{J}^{(\alpha)}, m_\alpha^2 \right]$ (which is the free generating functional of a single inhomogeneous field of mass $m_\alpha$, normalized to $\tilde{Z}_\alpha[0]=1$), with $\alpha = 1,2$, and  
\begin{equation}
 \langle \dots \rangle_{\bar{0}}^{\vec{J}_0} \equiv \frac{\Biggl\langle \sqrt{ \det \tilde{G_1}}^{N-1} \sqrt{ \det \tilde{G}_2} \dots \Biggr\rangle_{0}^{\vec{J}_0}}{\Biggl\langle \sqrt{ \det \tilde{G}_1}^{N-1} \sqrt{ \det \tilde{G}_2} \Biggr\rangle_{0}}. \label{zero-mode-exp-value}
\end{equation}
The normalization $\mathcal{N}$ is chosen to render $\mathcal Z[0] = 1$. The superindex $\vec{J}_0$ indicates that the $\bar{0}$-expectation value is taken over the zero modes in the presence of an external source $\vec{J}_0$. 
 
If the interaction terms in $S^{(3)}_{int}$  are  treated perturbatively,   a new type of perturbative corrections  introduced by 
\cite{LMT18} can be computed using new Feynman rules  
The  important differences from traditional Feynman rules are:  As a consequence of the definition of the new ``free'' generating functional in Eq.~(\ref{Z-LO})
as a weighted average over the zero modes, there is no direct cancellation of disconnected graphs when computing perturbative corrections. Indeed, consider a correction $\Delta \tilde{Z}$ to the generating functional of the inhomogeneous modes, which at LO is just $\tilde{Z}_1 \tilde{Z}_2$, prior to the zero mode average. The corrected complete (including both zero and inhomogeneous modes)   generating functional $\mathcal{Z}'$ now reads
\begin{equation}
 \mathcal{Z}'[\vec{J}_0, \tilde{J}^{(1)}_i, \tilde{J}^{(2)}] = \Bigl\langle \tilde{Z}_1[\tilde{J}^{(1)}_i] \,  \tilde{Z}_2[\tilde{J}^{(2)}] + \Delta \tilde{Z}[\tilde{J}^{(1)}_i,\tilde{J}^{(2)}] \Bigr\rangle_{\bar{0}}^{\vec{J}_0}, 
\label{Z-NLO}
\end{equation}
which corrects Eq.~(\ref{Z-LO}). Now consider a corrected n-point function of inhomogeneous fields computed from the previous  expression, treating  $\Delta \tilde{Z}$ perturbatively, 
\begin{eqnarray}
&&  \frac{1}{\mathcal{Z}'} \frac{\delta^n \mathcal{Z}'}{\delta \tilde{J}_{a_1}(x_1) \dots \delta \tilde{J}_{a_n}(x_n)} \Bigg|_{J=0}  =   \Biggl\langle \frac{\delta^n (\tilde{Z}_1 \tilde{Z}_2)}{\delta \tilde{J}_{a_1}(x_1) \dots \delta \tilde{J}_{a_n}(x_n)} \Bigg|_{\tilde{J}=0} \Biggr\rangle_{\bar{0}} \\
 &+& \Biggl\langle \frac{\delta^n \Delta \tilde{Z}}{\delta \tilde{J}_{a_1}(x_1) \dots \delta \tilde{J}_{a_n}(x_n)} \Bigg|_{\tilde{J}=0} \Biggr\rangle_{\bar{0}}  
 - \Bigl\langle \Delta \tilde{Z} \Bigr\rangle_{\bar{0}} \Biggl\langle \frac{\delta^n (\tilde{Z}_1 \tilde{Z}_2)}{\delta \tilde{J}_{a_1}(x_1) \dots \delta \tilde{J}_{a_n}(x_n)} \Bigg|_{J=0} \Biggr\rangle_{\bar{0}}\,,   \nn   \label{n-pt-NLO}
\end{eqnarray}
where we used the normalization $\tilde{Z}_\alpha[0]=1$. The first term on the right-hand side is the leading contribution obtained from Eq.~(\ref{Z-LO}), while the second and third terms are the corrections. In the usual case, the second term contains both connected and disconnected contributions, the latter of which are cancelled by the third term. However, in the current situation this does not occur due to the weighting over the zero modes.  Indeed, 
\begin{equation}
 \Biggl\langle \Delta \tilde{Z} \frac{\delta^n (\tilde{Z}_1 \tilde{Z}_2)}{\delta \tilde{J}_{a_1}(x_1) \dots \delta \tilde{J}_{a_n}(x_n)} \Bigg|_{J=0} \Biggr\rangle_{\bar{0}} \neq \Bigl\langle \Delta \tilde{Z} \Bigr\rangle_{\bar{0}} \Biggl\langle \frac{\delta^n (\tilde{Z}_1 \tilde{Z}_2)}{\delta \tilde{J}_{a_1}(x_1) \dots \delta \tilde{J}_{a_n}(x_n)} \Bigg|_{J=0} \Biggr\rangle_{\bar{0}}. 
\end{equation}
By adding and subtracting the left-hand side of the above equation to Eq.~(\ref{n-pt-NLO}), we can identify two contributions to the correction of the n-point function as follows: a connected part
\begin{eqnarray}
 \Delta \langle \tilde{\phi}_{a_1}(x_1) \dots \tilde{\phi}_{a_n}(x_n) \rangle_{C}&=& \Biggl\langle \frac{\delta^n \Delta \tilde{Z}}{\delta \tilde{J}_{a_1}(x_1) \dots \delta \tilde{J}_{a_n}(x_n)} \Bigg|_{\tilde{J}=0}   \nn\\
 &-& \Delta \tilde{Z} \frac{\delta^n (\tilde{Z}_1 \tilde{Z}_2)}{\delta \tilde{J}_{a_1}(x_1) \dots \delta \tilde{J}_{a_n}(x_n)} \Bigg|_{J=0} \Biggr\rangle_{\bar{0}}\,, \label{n-pt-C}
\end{eqnarray}
which is built in the standard way with the connected Feynman diagrams using the new rules; and a 0-connected part
\begin{eqnarray}
\Delta \langle \tilde{\phi}_{a_1}(x_1) \dots \tilde{\phi}_{a_n}(x_n) \rangle_{^{0} C} &=& \Biggl\langle \Delta \tilde{Z} \frac{\delta^n (\tilde{Z}_1 \tilde{Z}_2)}{\delta \tilde{J}_{a_1}(x_1) \dots \delta \tilde{J}_{a_n}(x_n)} \Bigg|_{J=0} \Biggr\rangle_{\bar{0}}  \nn\\
 &-&\Bigl\langle \Delta \tilde{Z} \Bigr\rangle_{\bar{0}} \Biggl\langle \frac{\delta^n (\tilde{Z}_1 \tilde{Z}_2)}{\delta \tilde{J}_{a_1}(x_1) \dots \delta \tilde{J}_{a_n}(x_n)} \Bigg|_{J=0} \Biggr\rangle_{\bar{0}},  
\end{eqnarray}
which accounts for contributions  that are  not connected by      $\tilde{G}_\alpha$, but  when written  in terms of the original perturbation theory  they are actually connected by  correlations of the zero-modes.
 
   A  systematic evaluation of the integrals  over the zero modes, $\langle \dots \rangle_{\bar{0}}$ can be implemented by  an  expansion  in    $1/N$ using  the saddle-point approximation (Laplace method) \cite{LMT18}. 

 \subsubsection{ Long wavelength behavior of the two-point functions for massless fields}\label{sec-all-N}
 
Once analytically continued back to LdS  we  can obtain the limiting value of the two-point functions for large distances/late times, $r\to \infty$. 
After the resummation of the bi-quadratic terms, the  two-point function (without any corrections from $S^{(3)}_{int}$) is 
  \begin{eqnarray}
  \langle \phi_a(x) \phi_b(x') \rangle &=& \frac{\delta_{ab}}{N} \Biggl\langle  |\vec{\phi}_0|^2 + (N-1) \tilde{G}_{1}(r) + \tilde{G}_{2}(r) \Biggr\rangle_{\bar{0}}. \label{full-resummed-2pt-f}
  \end{eqnarray}
Combining the constant contributions, we get
\begin{eqnarray}\label{constant-resummed}
 \langle \phi_a(x) \phi_b(x') \rangle^{LO} &\to& \frac{\delta_{ab}}{N} \Biggl\langle \frac{2N}{\lambda} u - \left(N-1+\frac{1}{3} \right) \frac{1}{V_d u} \Biggr\rangle_{\bar{0}} \nn\\
 &\simeq& -\sqrt{\frac{2}{V_d \lambda}} \frac{4\delta_{ab}}{3N} + \mathcal{O}(N^{-2}, \lambda^0),
\end{eqnarray}
where the expansion is at NLO in $1/N$.  
The reason why starting at this order we encounter a nonvanishing limiting value  is that at NLO in $1/N$ there are diagrams that contribute at LO in $\sqrt{\lambda}$ that are not of the type included in the resummation of the bi-quadratic terms \footnote{The family of diagrams that contribute at large-$N$ is that of the daisy and superdaisy type, which add a local part to the self-energy, and whose leading contribution in $\sqrt{\lambda}$ is already completely taken into account by the resummation.}.    This stems from the fact that, although in EdS there is a hierarchy between  interactions with $\vec{\phi}_0$ and $\vec{\tilde{\phi}}$ in terms of powers of $\sqrt{\lambda}$, upon analytic  continuation to LdS, when $r \to \infty$ the correlators of inhomogeneous modes receive an enhancement that makes them as relevant as the zero mode correlators,
\begin{equation}\label{newcounting}
 \langle |\vec{\phi_{0}}|^2 \rangle_{\bar{0}} \sim \frac{1}{\sqrt{\lambda}}, \quad\quad \langle \tilde{\phi}_a(x) \tilde{\phi}_b(x') \rangle_{\bar{0}} \to\sim \frac{\delta_{ab}}{\sqrt{\lambda}}.
\end{equation}
Therefore,  a consistent calculation of the two-point function at large distances  in LdS,   at  the lowest order (LO)  in $\sqrt{\lambda}$   and next-to-leading order (NLO)  in $1/N$,  requires further non-perturbative resummation.  In  \cite{LMT18} an additional  resummation, consistent  only at  LO in  $\sqrt{\lambda}$, but up to NLO in $1/N$ has been performed explicitly, with  results showing that the two point functions vanish  as $r\to +\infty$ also at  NLO   in $1/N$. 
The strategy implemented to perform such calculation exploits the fact that the  building blocks of the new Feynman rules involve the   propagators  $\tilde{G_1}$ and $\tilde{G_2}$  which are massive  (with $|\vec{\phi}_0|$-dependent masses). This allows one to  use     general
 theorems  proved in Refs.~\cite{MarMor10,MarMor11} for massive propagators. 
 
 To understand the asymptotic IR behaviour of the two-point functions, the next step is to study the decay of the full two-point function at large distances.  In principle the  same strategy can be applied systematically for a given order in $1/N$. This would  involve the evaluation of the most relevant $r$-dependent parts of all the diagrams that contribute at that order.   
Such a calculation has not been done so far.  What L\'opez Nacir,  Mazzitelli and Trombetta  presented  in \cite{LMT18}  are  the results  for  the asymptotic  behaviour  obtained   after resumming only the bi-quadratic terms (with  the exact treatment of the zero modes). The asymptotic scalings with $r$  are  summarized in Table~\ref{comparison-decays} for different values of $N$,  where also the standard perturbative results (prior to the resummation of bi-quadratic interactions) are shown for comparison.   
{\renewcommand{\arraystretch}{1.5}
\begin{table}
\begin{center}
\begin{tabular}{| c | c | c | c |}\hline
   & Free & $L$-loop & Resummed \\\hline
  $N=1$ & & & $\sqrt{\log(r)}$ \\ \cline{1-1} \cline{4-4}  
  $N=2$ & & & $\log(\log(r))$ \\ \cline{1-1} \cline{4-4}
  $N>2$ & $\log(r)$ &  $\log(r)^L$ & $\log(r)^{-(N-2)/2}$ \\ \cline{1-1} \cline{4-4} 
  $N \to \infty$ & & &  $r^{-m_{\rm dyn}^2/(d-1)}$  \\ \hline
\end{tabular} 
\end{center}
\caption{{\rm From \cite{LMT18}} Asymptotic behavior of the resummed two-point function in the  massless case ($m=0$) for different values of the number of fields $N$ as the de Sitter invariant distance $r \to +\infty$. After resummation of the bi-quadratic interaction terms, the usual logarithmic divergences present in the perturbative calculation (at all $N$) are softened when $N=1,2$ and cured for $N>2$. Here $m_{\rm dyn}^2=\sqrt{\lambda/(2V_d)}.$ }
\label{comparison-decays}
\end{table}  

In conclusion, the results obtained by \cite{LMT18} summarized in Table~\ref{comparison-decays} show that when the bi-quadratic interactions between the zero and higher modes are treated nonperturbatively,   the resummed two-point functions become  convergent or less divergent in the  large-distance limit, depending on the value of $N$.  
 The improved behavior comes from the fact that they
can be written as  weighted averages of free, massive two-point functions. This was anticipated  in  \cite{Hol12} for $N=1$. 
It is unclear whether additional corrections coming from the remaining   diagrams that contribute at each order in $1/N$, when treated nonperturbatively, might change this behavior. This can be particularly important for small $N$, for which the decay given by the current resummation is   milder, or absent altogether. For instance, if the higher modes were to have an additional dynamical mass coming from the interactions in $S_{int}^{(3)}$, this mechanism could dominate the IR behavior.


\section{Nonperturbative RG. Graviton and gauge issues}  

In this last section we mention three topics on this subject:   i) the work of Guillieux and Serreau \cite{GuiSer15} using nonperturbative renormalization group techniques; ii) the work of Moreau and Serreau \cite{MorSer18} on the backreaction of  superhorizon fluctuations of a light quantum scalar field on a classical de Sitter geometry by means of the Wilsonian renormalisation group;   iii) the  IR behavior of gravitons, where the gauge issue makes it more challenging.  This last topic continues into the last chapters of the book.

\subsection{Nonperturbative renormalization group}



Nonperturbative renormalization group (NPRG) methods originally formulated in \cite{BerTetWet02,Del12} are useful for infrared physics from critical phenomena to long distance dynamics of non-Abelian gauge fields. They have been adopted to de Sitter space-time \cite{Kay13,Ser14} for the study of  renormalization group (RG) flow of $O(N)$ scalar field theories at superhorizon scales. Gravitationally enhanced infrared fluctuations renders the RG flow to be effectively dimensionally reduced to that of a zero-dimensional Euclidean field theory. This has various consequences, such as  the radiative restoration of spontaneously broken symmetries in any space-time dimension.  The phenomenon of effective dimensional reduction allows one to establish a direct relation between the NPRG approach and the stochastic effective theory of Starobinsky and Yokoyama \cite{StaYok94}. Guillieux and Serreau \cite{GuiSer15} show  that the effective zero-dimensional field theory which results from integrating out the superhorizon degrees of freedom is equivalent to the late-time equilibrium state in the stochastic description.  
They also showed that the dimensionally reduced theory in (Lorentzian) de Sitter space-time at superhorizon scales is equivalent to the effective theory for the zero mode on the compact Euclidean de Sitter space \cite{HuOC87,Raj10,BenMoc13}.

For an $O(N)$ scalar field $\phi$ with a potential $V(\phi)$ being a function of the invariant $\phi_a \phi^a $ in an expanding flat-RW de Sitter  these authors derived a flow equation for the  beta function of the effective potential $\b (V'', \k)$ as a function of the curvature of the potential (a prime dentoes $d/d \phi$) or, alternatively, the order of the Bessel mode function $\n = \sqrt {(d/2)^2-V''}$ (\underline{Note}: the $d$ used here refers to spatial dimensionality while in earlier sections $d$ denotes spacetime dimensionality).
Both $\n$, $V''$ run with $\kappa$, an energy scale, with $\k >1 $ moving towards the ultraviolet and $\k =0$ where all quantum fluctuations are integrated out.  ($\ln \k$ is often called the RG `time' in the flow equation.) Comparing it with the beta function of the same theory in Minkowski space  they showed the de Sitter beta function coincides with the Minkowski one for all values of $V''$ in the regime of subhorizon scales $\k\gg 1$ and for all values of $\k$  when $V''\gg 1$. Spacetime curvature effects become sizable on superhorizon scales  $\k \gg 1$ for $V''_\k \approx d^2/4$.  
In a plot of the beta function $\b (V''; \k)$ as a function of $\ln \k$ for different values of the potential curvature $V''$ [see their Fig. 2] its slope is dramatically reduced and even turns to zero for $V'' \ll \k^2 \ll 1$ as a result of the gravitationally induced amplification of infrared fluctuations. This is a RG manifestation of the phenomenon of effective infrared dimensional reduction we discussed in the first part of this chapter.  The authors also derived the RG flow in a $ d+1$ dimensional sphere, the Euclidean de Sitter space.   
While the analysis of \cite{Ser14} was restricted to the deep infrared regime, where the flow is already dimensionally reduced, the analysis of \cite{GuiSer15} considered the complete flow from sub-horizon to super-horizon scales.  This makes possible the study of how a possible broken phase in the Minkowski regime gets restored once gravitational effects become important in the infrared regime.

\subsubsection{Recovery of stochastic results from NPRG flow}

It is instructive to see how  the Euclidean zero mode and infrared dimensional reduction, the themes of the first part of this chapter, can be   connected to the stochastic approach via the nonperturbative RG method. 
Infrared dimensional reduction suggests that the solution of the flow equation governed by the beta function 
can be written as an effective zero-dimensional field theory. Consider the following ordinary integral   \cite{GuiSer15}
\begin{equation}
\label{eq:ordinaryint}
e^{-\Omega_{d+1}{\cal W}_\kappa(J)} {=} {\int} d^N \phi \,e^{ -\Omega_{d+1}\left[V_{\rm eff}(\phi)  + J_a\phi_a +\frac{\kappa^2}{2}\phi_a\phi_a\right]},
\end{equation}
where $\Omega_{d+1}=4\pi^{d/2+1}/[d\Gamma(d/2)]$ and $V_{\rm eff}(\phi)$ is a function to be specified below. 
Introduce the Legendre transform 
\beq
\label{eq:legendre}
 V_\kappa(\phi)={\cal W}_\kappa(J)-J_a\phi_a-\frac{\kappa^2}{2}\phi_a\phi_a,
\eeq
with $\partial{\cal W}_\kappa(J)/\partial J_a=\phi_a$. It satisfies the flow equation 
for the case of small potential curvature $ |V''| \ll 1$ . 
One can adjust the function $V_{\rm eff}(\phi)$ so as to produce the appropriate initial conditions\footnote{In the case $N=1$, one can show that $V_{\rm eff}(\phi)\approx V_{\kappa_0}(\phi)$ if $V_{\rm eff}''(\phi)\ll\kappa_0^2$. For arbitrary $N$, the inequality should be satisfied by the largest eigenvalue of the curvature matrix $\partial^2V_{\rm eff}(\phi)/\partial\phi_a\partial\phi_b$.} for the infrared flow at a scale $\kappa_0\sim 1$. All solutions of the flow equation in the deep de Sitter regime can thus be written as \eqref{eq:ordinaryint}. Notice that in this regime the original $D$-dimensional Lorentzian theory, with complex weight $\exp(iS)$ eventually flows to a zero-dimensional Euclidean-like integral, with real weight $\exp(-\Omega_{D+1}V_{\rm eff})$. Its physical meaning is ingrained in the dynamical finite size effect we discussed earlier. 

Let us now examine the \textit{stochastic approach} of Starobinsky and Yokoyama ~\cite{StaYok94} in this light.  Infrared dimensional reduction also holds the key to the  utility of the effective theory for fields of long wavelength modes on superhorizon scales.  Since these superhorizon modes, called $\bar \phi$ before, are almost frozen in time -- recall  an exponential expansion amounts to a scaling transformation in the effectively static picture -- they can essentially be described by a single degree of freedom,  a stochastic variable 
$\varphi_a(t)$ in each direction in field space, with $t$ the cosmological time.  These  variables become stochastic because we have added on them the influence of the subhorizon modes of shorter wavelengths,  represented by white noise in Starobinsky's model  \cite{Sta86}. They obey the by-now familair Langevin equation 
\beq\label{eq:Langevin}
 \partial_t\varphi_a(t)+\frac{1}{d}\frac{\partial V_{\rm soft}(\varphi)}{\partial\varphi_a(t)}=\xi_a(t),
\eeq
where $V_{\rm soft}(\varphi)$ is the potential seen by the long wavelength modes.
Treating the short wavelength modes as noninteracting fields in the Bunch-Davies vacuum, one has, generalizing the calculation of \cite{StaYok94,BenMoc13} to arbitrary $N$,
\beq
 \langle\xi_a(t)\xi_b(t')\rangle=\frac{\Gamma(d/2)}{2\pi^{{d\over2}+1}}\delta_{ab}\delta(t-t').
\eeq
Using standard manipulations, \eqref{eq:Langevin} can be turned into the following Fokker-Planck equation for the probability distribution ${\cal P}(\varphi,t)$ of the stochastic process
\beq
\label{eq:FP}
 \partial_t{\cal P}=\frac{1}{d}\frac{\partial}{\partial\varphi_a}\left\{\frac{\partial V_{\rm soft}}{\partial\varphi_a}{\cal P}+\frac{1}{\Omega_{D+1}}\frac{\partial{\cal P}}{\partial\varphi_a}\right\}.
\eeq
The latter admits an $O(N)$-symmetric stationary attractor solution at late times (i.e., in the deep infrared), given by
\beq
 {\cal P}(\varphi)\propto\exp\big\{-\Omega_{D+1}V_{\rm soft}(\varphi)\big\}.
\eeq
Equal-time correlation functions on superhorizon scales can then be computed as moments of this distribution. This is the point of contact one can make between the  stochastic approach and the RG analysis described above, namely,  \eqref{eq:ordinaryint}  in the limit $\kappa_0$ provided one identifies $V_{\rm soft}(\varphi)=V_{\rm eff}(\varphi)\approx V_{\kappa_0}(\varphi)$. For instance, one has
\beq
 \langle\varphi_a\varphi_b\rangle=\frac{\displaystyle\int d^N\varphi\;\varphi_a\varphi_b\,{\cal P}(\varphi)}{\displaystyle\int d^N\varphi\;{\cal P}(\varphi)}=\left.\frac{1}{\Omega_{D+1}}\frac{\partial^2{\cal W}_{\kappa=0}(J)}{\partial J_a\partial J_b}\right|_{J=0}.
\eeq
Note that the relevant potential used in the stochastic approach is  not the one at the UV scale $\Lambda$ but the one evolved down to the horizon scale $\kappa \to 0$, a clear manifestation that this theory is engineered for the infrared regime.

\subsubsection{Going beyond the local effective potential by derivative expansion }  
As was mentioned in the beginning of this chapter, to study the symmetry behavior of quantum fields in a dynamical spacetime one needs to take into consideration the time-varying background field. The effective potential which assumes a constant background field $\bar \phi$ is only the zeroth order approximation in a derivative expansion of the background field,  the first order capturing  slow variations. This was discussed in Chapter 2 under the topic of quasi-local effective Lagrangian following \cite{HuOC84}.  Now for the IR problem in de Sitter one would expect that the late time behavior of the inflaton may not be so sensitive to this correction because the modes have virtually all been red-shifted exponentially fast, and the differences arising from the time variation of the background field may show up as a small logarithmic correction. This aspect has been addressed by Guillieu and Serreau \cite{GuiSer17} who 
introduced a so-called `local potential approximation prime' which includes a running (but field-independent) field renormalization to the full effective potential. They explicitly computed the associated anomalous dimension for $O(N)$ theories and find that it can take on large values along the flow, leading to sizable differences as compared to the local potential approximation. However, it does not prevent the phenomenon of gravitationally-induced dimensional reduction. As a consequence, the effective potential at the end of the flow is unchanged as compared to the local potential approximation, the main effect of the running anomalous dimension being merely to slow down the flow. 

\subsubsection{Stability of dS against backreaction of superhorizon quantum fluctuations}
We close this topic by mentioning the recent work of Moreau and Serreau \cite{MorSer18} who  have investigated the backreaction of  a light quantum scalar field on a de Sitter geometry by means of NPRG techniques. They found a nontrivial renormalisation of the spacetime curvature as superhorizon fluctutations are progressively integrated out. Perturbative loop corrections grow unbounded as a 
result of the gravitational amplification of such fluctuations. This signals the breakdown of perturbation theory rather than an instability. Nonperturbative effects
come into play with, in particular, the dynamical generation of a mass, which screens the growth of superhorizon fluctuations and freezes the RG flow of the effective
spacetime curvature. Overall, the infrared renormalisation of the latter is controlled by the gravitational coupling which is small by assumption in their semiclassical treatment. This work adds weight to the belief that  de Sitter spacetime  is stable against infrared quantum fluctuations.

\subsection{IR divergence in graviton and the gauge issue}
This is an important and challenging subject. We  highlight some  key issues involved  without belaboring,  but  point to the sources where one can find more details of the topics covered. 

\subsubsection{Graviton's IR behavior}

Quantized linear gravitational perturbations,  otherwise known as gravitons,  have two polarizations, each obeying an equation of motion identical to that of a massless minimally coupled scalar field \cite{Gri74,ForPar77}. For this reason one may consider  their infrared behavior to be identical.  However, there are significant differences between a scalar field  and the gravitational perturbations 
\cite{FroRouVer12}.  For one, the (partial) derivative character of the gravitational interaction improves  the IR behavior over that of massless minimally coupled matter fields.  A more distinct difference is the need to consider appropriate gauge-invariant observables for gravitons, which is a rather nontrivial aspect even in perturbative quantum gravity.
Moreover, one should restrict one's attention to ``sufficiently local" observables \cite{GidMarHar06}  that properly characterize the geometrical properties within a region of finite physical size.  This is a crucial factor for the  construction of IR-safe observables in situations which would otherwise lead to divergences in the absence of an IR cut-off \cite{UraTan10,SenZal10}. 

\paragraph{The gauge issue}  
 It is essential to distinguish whether these IR divergences are restricted to the gauge sector of linearized gravity, or they appear also in the physical sector.  Allen \cite{All87} and Higuchi \cite{Hig87} have shown that in the traceless-transverse-synchronous gauge the IR-divergent part of the graviton two-point function can be expressed in a nonlocal pure-gauge form. Higuchi,  Marolf and Morrison \cite{HigMarMor11b} observed that a local gauge transformation on the graviton modes is sufficient to eliminate the IR divergences plaguing the graviton two-point function in that gauge.  Other gauges and coordinate systems in which  the graviton two-point function is IR finite have been worked out. See, e.g.,  references cited in \cite{FroHigLim16}.

 Another way to examine the gauge nature of these IR divergences is through the linearized Weyl tensor, which is a local and gauge-invariant observable in the linearized theory. It was shown that the two-point function of the linearized Weyl tensor is IR finite even if computed using a de Sitter noninvariant graviton two-point function with an IR cutoff \cite{MorTsaWoo12b,FroRouVer14}. We will describe this approach in more detail in a later chapter.  
 
\paragraph{Residual Gauge conditions} 
In cosmological perturbation theory,  one usually chooses gauge conditions on the entire constant time slice. But from an observational perspective one can only see within a causally connected region which changes in time. 
To regularize the IR contributions for the curvature perturbation, it is necessary to take into account this subtle issue.  (See the series of papers by Tanaka and Yurakawa cited in \cite{TanUra14}).

Since only physical quantities in the observable region are of relevance,   one should compute observable quantities unaffected by what is outside of the observable region.   The choice of gauge conditions should respect the compatibility of the `outside' degrees of freedom with what is observed.  These so-called  `residual' gauge conditions can be linked  to the degrees of freedom in the boundary conditions of our observable local universe. It was shown that requesting the invariance under the change of such residual coordinate degrees of freedom in the local universe can ensure the IR regularity and the absence of the secular growth. The otherwise singular IR contributions are subsumed in the residual coordinate degrees of freedom \cite{TanUra13}.  On the issue whether the residual coordinate degrees of freedom can also affect the IR behavior of the graviton  Tanaka and Urakawa \cite{TanUra14}  showed that when invariance under the residual coordinate transformations are required, the IR regularity and the absence of secular growth are also guaranteed for the graviton loops \footnote{This is contested by Tsamis and Woodard and co-workers  who have done extensive perturbative calculations on two loop graviton contributions.  They maintain that the secular behavior is not a  gauge effect but is physical and can have important consequences such as causing the decay of the cosmological constant.  See, e.g., the exchanges between Tsamis \& Woodard \cite{TsaWoo08} on the one side  vesus Garriga \& Tanaka on the other \cite{GarTan08}.}. 

\paragraph{Nonlinear effects} 
Measurement of primordial gravitational waves can provide information which cannot be obtained by the measurement of scalar perturbations. It could also provide finer discrimination between different models of inflation.  Unlike the amplitude of the adiabatic curvature perturbation, the amplitude of the gravitational waves is  sensitive to the detailed dynamics of the inflationary universe, and thus  provide the energy scale of inflation. Measurements in the primordial gravitational waves for non-linear perturbations \cite{LosUnr05,LosUnr08}  or interactions  \cite{MalPim11,SodKod11} can for example reveal the impact of parity violation in the gravity sector on the bi-spectrum of the primordial gravitational waves.   

\subsubsection{Graviton loop corrections}
We continue with the narrative in  \cite{TanUra14} on this issue. 
Since a massless scalar field yields a scale-invariant spectrum in the IR limit as $P(k) \propto  1/k^3$, a naive loop integral yields a factor $\int d^3k/k^3 \approx  \int k/k$, which diverges logarithmically.  Free gravitons are expected to have the same IR behavior, as each of the two polarizations behaves like a massless minimally coupled scalar field.  As the graviton loop corrections show divergences,   regularization scheme is needed. A simple minded way is to introduce an IR cutoff, say, at a
comoving scale $k_{\textsc{ir}}$.  However, this will not provide a satisfactory solution, because the loop integral of the super-Hubble modes gives $\int^{aH}_ {k_{\textsc{ir}}} dk/k  \approx \ln(aH/k_{\textsc{ir}}) $, which logarithmically increases in time. (Here $a$ is the scale factor and $H$ is the Hubble parameter of the background spacetime.)  Compared with the tree-level contribution, the loop corrections are typically suppressed by the Planck scale as $(H/M_P)^2$ with $M^2 _P \equiv (8\pi G_{\textsc{n}})^{-1}$.

Fr\"ob, Roura and Verdaguer \cite{FroRouVer12} calculated the one-loop correction to the tensorial metric perturbations,  the key ingredient to obtain the `sufficiently local' observables \cite{GidMarHar06}.  (The scalar and vectorial metric perturbations can be directly obtained from the stress tensor correlation function.)  By employing  a large $N$ expansion for $N$ matter fields interacting with the gravitational field \cite{HuRouVer04a}  they managed to avoid the consideration of graviton internal loops (the lowest-order contributions to the connected two-point function of the metric perturbations are of order $1/N$,  whereas any contribution including graviton loops is suppressed by higher powers of $1/N$). This issue will be taken up in a later chapter.  \\

\noindent {\bf Acknowledgments} It is with great pleasure that I express my long-held appreciation to Denjoe O'Connor, whose pioneering work in his Ph.D. thesis \cite{OCPhD}  enhanced my understanding of this fundamental subject immensely. I thank Richard Woodard for sharing his joy in discovering the ``miracle" in the work of Starobinksy with Yokoyama. Kudos to Alexei, my contemporary in the crack-of-dawn days of quantum field theory in curved spacetime, for his life-long achievements of eminence. Respects to Cliff Burgess, Jaume Garriga, Atsushi Higuchi, Stefan Hollands, Don Marolf,  Julien Serreau, Takahiro Tanaka  and the recent authors of nonperturbative RG for advancing this rich old field to higher  levels and new frontiers. I thank  Diana L\'opez Nacir and Diego Mazzitelli for discussions on their recent work on Hartree approximation and on the infrared behavior of Lorentzian quantum field theory and for their careful reading of this manuscript catching many typos. Same to Jen-Tsung Hsiang  for checking the signs and conventions with meticulous care. Last but not least, I take this opportunity to express my warm appreciation to my co-author, Enric Verdaguer, for  his high standards and impeccable scholarship.    

\bibliography{refs18d-}\label{refs}
 \bibliographystyle{unsrtnat}


\end{document}